\documentstyle[psfig,12pt]{article}
\def\TL{\hfil$\displaystyle{##}$}
\def\TR{$\displaystyle{{}##}$\hfil}

\def\TT{\hbox{##}}
\def\seqalign#1#2{\vcenter{\openup1\jot
  \halign{\strut #1\cr #2 \cr}}}

\def\comment#1{}
\def\fixit#1{}

\def\mop#1{\mathop{\rm #1}\nolimits}

\def\diag{\mop{diag}}
\def\tr{\mop{tr}}

\def\overleftrightarrow#1{\vbox{\ialign{##\crcr
     $\leftrightarrow$\crcr\noalign{\kern-0pt\nointerlineskip}
     $\hfil\displaystyle{#1}\hfil$\crcr}}}

\def\lsim{\mathrel{\mathstrut\smash{\ooalign{\raise2.5pt\hbox{$<$}\cr\lower2.5pt\hbox{$\sim$}}}}}
\def\gsim{\mathrel{\mathstrut\smash{\ooalign{\raise2.5pt\hbox{$>$}\cr\lower2.5pt\hbox{$\sim$}}}}}

\def\sqr#1#2{{\vcenter{\vbox{\hrule height.#2pt
         \hbox{\vrule width.#2pt height#1pt \kern#1pt
            \vrule width.#2pt}
         \hrule height.#2pt}}}}
\def\square{\mathop{\mathchoice\sqr56\sqr56\sqr{3.75}4\sqr34\,}\nolimits}

\def\href#1#2{#2}  

\def\lbldef#1#2{\expandafter\gdef\csname #1\endcsname {#2}}
\def\eqn#1#2{\lbldef{#1}{(\ref{#1})}%
\begin{equation} #2 \label{#1} \end{equation}}
\def\eqalign#1{\vcenter{\openup1\jot
    \halign{\strut\span\TL & \span\TR\cr #1 \cr
   }}}
\def\eno#1{(\ref{#1})}

\def\vphi{{\vec\varphi}}
\def\comment#1{  \begin{raggedright}{\tt [#1]}\end{raggedright}}
\begin{document}
\baselineskip=15.5pt
\pagestyle{plain}
\setcounter{page}{1}

\begin{titlepage}

\begin{flushright}
PUPT-1916 \\
hep-th/0002160
\end{flushright}
\vfil

\begin{center}
{\huge Curvature Singularities:}
\vskip0.5cm
{\huge the Good, the Bad, and the Naked}
\end{center}

\vfil
\begin{center}
{\large Steven S. Gubser}
\end{center}

$$\seqalign{\span\TL & \span\TT}{
& Joseph Henry Laboratories, Princeton University, Princeton,
NJ 08544
}$$
\vfil

\begin{center}
{\large Abstract}
\end{center}

\noindent 
 Necessary conditions are proposed for the admissibility of singular
classical solutions with $3+1$-dimensional Poincar\'e invariance to
five-dimensional gravity coupled to scalars.  Finite temperature
considerations and examples from AdS/CFT support the conjecture that
the scalar potential must remain bounded above for a solution to be
physical.  Having imposed some restrictions on naked singularities
allows us to comment on a recent proposal for solving the cosmological
constant problem.

\vfil
\begin{flushleft}
February 2000
\end{flushleft}
\end{titlepage}
\newpage
\section{Introduction}
\label{Introduction}

In AdS/CFT \cite{juanAdS,gkPol,witHolOne} (see \cite{MAGOO} for a
review), bulk geometries which are only asymptotically $AdS_5$ near
the timelike boundary are dual to relevant deformations of the CFT or
to non-conformal vacua.  Far from the timelike boundary, various
singularities might arise.  There must be some restrictions on the
type of singularity which is allowed: for instance, we take it as
obvious that negative mass Schwarzschild should be excluded on the
grounds of vacuum stability \cite{hmValue}.  The goal of this paper is
to address the question of what singularities are allowed in
geometries of the form
  \eqn{AnsatzOne}{\eqalign{
   ds^2 &= e^{2 A(r)} (-dt^2 + d\vec{x}^2) + dr^2  \cr
   \vphi &= \vphi(r) \,,
  }}
 where by assumption $A(r) \sim r/L$ and $\vphi(r) \sim (const)$ for
large $r$.  $L$ is the radius of curvature of the asymptotically
$AdS_5$ region.  Geometries of the form \AnsatzOne\ arise in the
gauged supergravity description of certain vacuum states on the
Coulomb branch of ${\cal N}=4$ super-Yang-Mills theory
\cite{klt,kwTwo,fgpwTwo,BrandhuberSfetsosOne,cglp}, and are also
suspected to be the supergravity duals of most renormalization group
flows arising from relevant deformations of the lagrangian of the CFT
\cite{kwOne,gns,lopez,gppzOne,dzOne,kpw,klm,fgpwOne,gppzTwo,gppzThree,
gStrings,vvOne,knLog}.

The classical equations of motion, plus either positivity of the
scalar kinetic terms or the more general ``null energy'' condition,
together imply $A''(r) \leq 0$ \cite{gppzOne,fgpwOne}.  Since we
assume $A(r) \sim r/L$ with $L > 0$ for large $r$, $A(r)$ must in fact
be monotonically increasing with $r$.  Thus a curvature singularity can
arise for the metric \AnsatzOne\ if $A(r) \to -\infty$ at some finite
$r_0$.\footnote{Singularities where $A(r) \to -\infty$ at finite $r_0$
are the only type that will be encountered in this paper, but for the
sake of completeness let us review three other possibilities.  First,
it could be that $A(r)$ is defined for all $r$, but $A'(r) \to \infty$
as $r \to -\infty$.  Supersymmetric examples with this behavior arise
in certain $U(1)$ gauged supergravities.  Second, an $S^5$ shell of
D3-branes produces a near-horizon geometry which is $AdS_5 \times S^5$
outside the shell and flat ${\bf R}^{10}$ inside the shell
\cite{ChepelevRoiban,GiddingsRoss}.  The shell itself produces a
curvature singularity at finite $A(r)$.  Similar singularities arise
in Horava-Witten theory compactified on a Calabi-Yau three-fold with
M5-branes in the bulk (see for example \cite{OvrutFive}): in the
five-dimensional description, there is a jump in $A'(r)$ across each
M5-brane.  This sort of singularity also arises in type~I$'$ with
D8-branes in the bulk.  Finally, there can be a codimension one
orientifold plane at finite $A(r)$ on which the five-dimensional
spacetime ends.  Again the examples are type~I$'$ and Horava-Witten
theory.}

The UV-IR relation \cite{SusskindWitten,PeetPolchinski} suggests that
curvature singularities that arise as $A(r) \to -\infty$ represent
non-trivial infrared physics in the dual field theory.  (Recall that
$A(r) = {1 \over 2} \log |g_{tt}|$ is roughly the log of the energy
scale).  This interpretation is natural from the point of view of the
holographic renormalization group \cite{vvOne}.  Suppose the field
theory vacuum is one which can support finite temperature: by this we
mean that the vacuum can be approached smoothly in a $T \to 0$ limit,
with no zero temperature phase transition.  Then a putative dual
description in supergravity should also be able to support finite
temperature, in the form of a black hole horizon which ``cloaks'' the
curvature singularity.  As the temperature is lowered to zero, the
horizon should retreat until the original singular geometry is
recovered.  Thus we arrive at a weak form of Cosmic Censorship: the
only curvature singularities allowed in geometries of the form
\AnsatzOne\ are those which can be obtained as limits of regular black
holes.

Even this statement is too strong: there are interesting vacua in
field theory which cannot support finite temperature.  Examples
include states on the Coulomb branch of ${\cal N}=4$ gauge theory.
The dual geometries are typically singular in five dimensions.  Any
finite temperature, no matter how small, would draw a Coulomb branch
state to the origin of moduli space, where conformal invariance is
recovered in the zero temperature limit.  Correspondingly, finite
temperature in the bulk should pull the dual singular geometries back
to AdS-Schwarzschild.  

As detectives in search of a useful rule for distinguishing good
singularities from bad, let us then ask the following two questions.
(1) Do the singular five-dimensional geometries corresponding to
Coulomb branch states share any common features with geometries which
are limits of regular black holes?  (2) Can we find other clear
examples which could help us establish the identifying marks of
healthy singularities as opposed to pathological ones?  The main goal
of this paper is to answer both questions affirmatively, and in fact
to argue that
  \eqn{Conjecture}{\eqalign{
   &\hbox{\sl Large curvatures in geometries of the form \AnsatzOne\ 
     are allowed only if}  \cr
   &\hbox{\sl the scalar potential is bounded above in the 
     solution.}
  }}
 This conjecture is peculiar in that it is the opposite of the naive
intuition from classical relativity.  The dominant energy condition,
for example, constrains the scalar potential to be non-negative---not
just in solutions, but in the lagrangian.  (This of course rules out
an $AdS_5$ solution, since the scalar potential is the cosmological
constant).  Nevertheless, it will be proven that \Conjecture\ is
indeed a necessary condition for a bulk geometry to support finite
temperature in the form of a black hole horizon.  It will also be
shown that \Conjecture\ correctly predicts some non-trivial aspects of
the vacuum structure of mass-deformed ${\cal N}=4$ gauge theory.

It is important to bear in mind that \Conjecture\ is a constraint on
{\it solutions}, not on the five-dimensional lagrangian.  In point of
fact, the scalar potential for $d=5$ ${\cal N}=8$ gauged supergravity
(from which all our examples will be drawn) is unbounded both above and
below.  There are solutions where scalars diverge near the singularity
in such a way that the scalar potential goes to $+\infty$.  These are
bad, according to \Conjecture.  There are others where the scalar
potential goes to $-\infty$.  These are good.  Thanks to AdS/CFT, we
can not only demonstrate that the good singularities have sensible
field theory duals, but in most cases even trace the sickness of the
bad singularities back to some definite pathology in the dual field
theory.

A complementary line of attack, which will not be pursued in this work
but is potentially very fruitful, is to resolve singularities in
metrics of the form \AnsatzOne\ by lifting them to ten dimensions and
finding a configuration of branes that produces the ten-dimensional
geometry (possibly in some near-horizon, large $N$ limit).  This is in
general a difficult task, but when it can be carried out, it provides
a satisfying account of how string theory resolves a singularity.  A
goal of the present paper is to try to anticipate which
five-dimensional singularities will admit a brane resolution.  The
methods should be equally applicable to Calabi-Yau compactifications
of Horava-Witten theory where one end-of-the-universe brane has
retreated to infinite redshift.  This is a case where our
understanding of the microscopic eleven-dimensional theory is so
incomplete that supergravity methods, combined with consistency
conditions such as anomaly inflow, may be the only tools available.

One may legitimately inquire why any method based on five-dimensional
supergravity should have predictive power regarding the nature of
singular solutions.  This is where AdS/CFT helps: confusing
singularities on the AdS side are often transparent in the field
theory dual.  In the end, however, the hope is that \Conjecture\ is a
robust result that will survive beyond the approximations in which we
have studied it, and even beyond AdS/CFT.  The nature of a local
singularity should be independent of the geometry far from it, so
having an asymptotically $AdS_5$ region may be just a crutch for
calculation.  The physics of the naked singularity should be as
independent of the asymptotically $AdS_5$ region as infrared effective
theories are of their detailed microscopic origins.

The organization of the paper is as follows.  Section~\ref{Background}
is a more technical introduction in which conventions are set and a
method for generating solutions is reviewed.  In
section~\ref{Motivation} we discuss the motivations for \Conjecture,
and we explain some related conditions which are useful when looking
at examples.  We also discuss the infrared asymptotics of generic
solutions and make some speculations regarding phase transitions.  In
section~\ref{Examples} we review three examples of asymptotically
$AdS_5$ geometries which arise as deformations of ${\cal N}=4$
super-Yang-Mills theory by a relevant operator.  In
section~\ref{Fluctuations} we consider linearized fluctuations around
a general asymptotically $AdS_5$ background and propose a
generalization of the Breitenlohner-Freedman bound.  Concluding
remarks and some further conjectures are presented in
section~\ref{Discussion}.  In the appendix we remark on brane-world
scenarios in which there is a curvature singularity parallel to our
world, at some finite proper distance from it.

\section{Background}
\label{Background}

All the geometries we will consider are classical solutions to the
equations of motion following from an action
  \eqn{WAction}{
   S = \int d^5 x \, \sqrt{g} \left[ {1 \over 4} R - 
    {1 \over 2} (\partial\vphi)^2 - V(\vphi) \right]  \,.
  }
 Usually it will be assumed that the potential $V(\vphi)$ has a local
maximum at $\vphi=0$.  Our main example of such an action is the
gravity-plus-scalars sector of $d=5$ ${\cal N}=8$ gauged supergravity
\cite{grw,ppn}.  Here the $42$ scalars $\vphi$ parametrize the coset
$E_{6(6)}/USp(8)$, and $V(\vphi)$ is a $SO(6) \times SL(2,{\bf
R})$-invariant function on this coset.  The kinetic term would be more
accurately represented as $G_{IJ}(\vphi) \partial\varphi^I
\partial\varphi^J$, where $G_{IJ}(\vphi)$ is the natural sigma model
metric for the coset.  In the general discussion that follows, we will
suppress factors of $G_{IJ}$ and its inverse.  Restoring them does not
make any difference to the essential points of the story as long as
$G_{IJ}$ is a smooth function of $\vphi$, which is true for $d=5$
${\cal N}=8$ gauged supergravity.  In the examples of
section~\ref{Examples} we will work on submanifolds of $E_{6(6)}$
where the metric is flat, so one may find a basis of scalars where
$G_{IJ} = \delta_{IJ}$.  We will call such scalars canonically
normalized.

The $40$ scalars that participate in $V(\vphi)$ are tachyonic at the
maximally supersymmetric point representing unperturbed ${\cal N}=4$
super-Yang-Mills theory.  In AdS/CFT they correspond to dimension~$2$
operators in the ${\bf 20}'$ of $SO(6)$ and dimension~$3$ operators in
the ${\bf 10} + \overline{\bf 10}$.  The remaining scalars are the
exactly massless dilaton and axion, corresponding to the exactly
marginal complex operator $\tr (F^2 + i F \tilde{F} + \ldots)$.  These
massless scalars will not participate in the geometries we consider.
Our remarks can be applied more generally to less supersymmetric
AdS/CFT duals, the key point being that the scalars in question are
tachyonic at $\vphi=0$.

It is often useful to take a potential of the form
  \eqn{VWForm}{
   V(\vphi) = {1 \over 8} \left( {\partial W \over \partial\vphi} \right)^2 - 
    {1 \over 3} W(\vphi)^2  \,,
  }
 where $W$ is some appropriate function of the scalars.  In $U(1)$
gaugings of $d=5$ ${\cal N}=2$ supergravity, $W$ can be identified
with the superpotential.  In \cite{kpw,fgpwOne,fgpwTwo,gppzThree} it
was shown that the potential of $d=5$ ${\cal N}=8$ gauged supergravity
also assumes the form \VWForm\ on restricted subspaces of the scalar
manifold, with $W$ identified as an eigenvalue of an $SO(6) \times
SL(2,{\bf R})$-invariant $USp(8)$ matrix $W_{ab}$, defined on
$E_{6(6)}/USp(8)$, which serves as a generalized superpotential in the
sense that the gravitino variation has the form $\delta\psi_{\mu a} =
{\cal D}_\mu \epsilon_a - {1 \over 6} g W_{ab} \gamma_\mu \epsilon^b$
(see \cite{grw,fgpwOne} for further details).  

In \cite{MyersUnpublished,SkenderisTownsend,dfgk}, the form \VWForm\
was found to be useful, independent of supersymmetry, for generating
solutions of the form \AnsatzOne\ and investigating their properties.
Indeed, given $V(\vphi)$, the equation \VWForm\ can be regarded as a
PDE for $W$, to be solved with arbitrary but specified initial
conditions.  It was observed in \cite{vvOne} that the $W$ so obtained
is related to Hamilton's principle function.  Of the possible forms
for $W$ satisfying \VWForm, only one (or at most discretely many) can
be related to the supersymmetry transformations.  However, if \VWForm\
can be solved systematically, then it is possible to generate any
solution of the form \AnsatzOne\ (supersymmetric or not) to the
equations of motion by solving the first order equations
  \eqn{FirstOrder}{
   \vphi' = {1 \over 2} {\partial W \over \partial\vphi} \qquad
   A' = -{1 \over 3} W(\vphi) \,.
  }
 Except in section~\ref{Fluctuations}, primes will indicate
derivatives with respect to $r$.  It can happen that $W$ is not a
single-valued function of $\vphi$, but rather a multi-sheeted graph in
the allowed region of $W$--$\vphi$ space where $V(\vphi) + {1 \over 3}
W(\vphi)^2 \geq 0$.  In such a case, \FirstOrder\ must be supplemented
with selection rules for going from one branch to another when the
boundary of the allowed region is reached (see \cite{dfgk} for a more
detailed discussion).  This subtlety will not be a concern in this
paper.

The {\it prima facia} evidence that the solution space of \VWForm\
together with \FirstOrder\ coincides with the solution space of the
equations of motion following from \WAction\ is that the number of
integration constants is the same.  In view of the realization
\cite{vvOne} that \FirstOrder\ can be recovered in Hamilton-Jacobi
theory, there can be no question that \VWForm\ together with
\FirstOrder\ are nothing more nor less than a fancy rewriting of the
equations of motion.  Nevertheless, it will be instructive to follow
through the counting of integration constants.  Suppose there are $n$
scalars.  The equations of motion consist of $n$ second order
differential equations, one for each scalar, plus one first order
constraint from the $G^r{}_r$ Einstein equation.  That means $2n+1$
integration constants.  In the first order formalism, $n$ of these
integrations constants are tucked into $W$ as constants of the motion;
$n$ more come from the first order equations for the scalars; and the
last comes from the equation for $A'$ in \FirstOrder, which is
equivalent to the $G^r{}_r$ Einstein equation if the first order
equations for the scalars hold.

There are various reasons to think that the space of solutions of the
form \AnsatzOne\ to the equations of motion following from \WAction\
is too big---that not all solutions are physical.  First, from a field
theory perspective, it is possible that the vacuum is ill-defined.
This would happen, for instance, if the effective potential in the
field theory were unbounded below.  It was suggested in \cite{dzOne}
that exactly this problem was responsible for the unphysical nature of
the $SO(5)$ critical point.  To flow to the $SO(5)$ point, it is
necessary to deform the ${\cal N}=4$ lagrangian by a term $m^2 \tr
(-X_1^2 - X_2^2 - X_3^2 - X_4^2 - X_5^2 + 5 X_6^2)$.  This means the
effective potential $V_{\rm eff}$ in the field theory is unbounded
below.  This pathology manifests itself in five-dimensional gauged
supergravity as a violation of the Breitenlohner-Freedman bound
\cite{BF,TownsendBF} at the $SO(5)$ critical point.  Another pathology
which one could imagine in field theory is attempting to assign a
negative vacuum expectation value (VEV) to a positive definite
operator.  In section~\ref{Examples} we will encounter examples where
this problem arises.

Generic solutions of the form \AnsatzOne\ to the equations of motion
following from \WAction\ have curvature singularities at finite $r$.
A very few, such as $AdS_5$ itself and the RG flows of
\cite{gppzOne,dzOne,fgpwOne}, have curvatures everywhere bounded.  We
believe we are on safe ground when we assert that not all naked
singularities are physical, even if we demand $3+1$-dimensional
Poincar\'e invariance and/or some unbroken supersymmetry.  Indeed,
some of the Coulomb branch flows explored in
\cite{fgpwTwo,BakasSfetsos,cglp} have an unphysical property when they
were lifted to ten dimensions: the brane distributions involve branes
both of positive tension and positive charge, and of negative tension
and negative charge.  A similar problem was beautifully resolved in
\cite{jpp}, but for Coulomb branch states it is obscure what
enhancement of symmetry could play a relevant role.  In
section~\ref{Examples} we will encounter other pathologies in
geometries which violate \Conjecture.  In short, it seems hard to make
a convincing case that all the nakedly singular, Poinar\'e-invariant
geometries that we can generate as solutions to the equations of
motion are physical.

We would however take the position that there are physically allowable
naked singularities in AdS/CFT, representing interesting infrared
physics in the dual field theory---only supergravity has limited
calculation power in their vicinity.  The repulson\slash
en\-han\-\c{c}on of \cite{jpp} is one example; the dual to ${\cal
N}=1$ super-Yang-Mills theory \cite{gppzThree} is another, which we
will have more to say about later.  In general, it is a non-trivial
exercise to lift five-dimensional solutions to ten dimensions and find
a brane configuration which realizes the geometry in some appropriate
limit.  And it is painful to proceed example-by-example, looking for
known pathologies, like negative VEV's for positive definite
operators, or imaginary ten-dimensional metrics.  Thus it is useful to
have a criterion such as \Conjecture, which is easy to apply and seems
fairly reliable.

\section{Motivating the criterion}
\label{Motivation}

In this section we will provide two motivations for \Conjecture.  The
first is that it is a necessary condition for the existence of
near-extremal generalizations of a solution to \FirstOrder.  We will
prove this assertion in section~\ref{FiniteT}.  By near-extremal
generalizations we mean a family of black hole solutions, with
horizons hiding the naked singularity, which converge to the
Poincar\'e invariant solution in an appropriate topology.  The Hawking
temperature of the horizon is identified with finite temperature in
the dual field theory.

Finite temperature in field theory serves as an infrared cutoff in the
sense that it masks physics at scales lower than the temperature.  A
naked singularity that can be hidden behind a black hole horizon is a
signal of non-trivial but sensible infrared physics in the dual field
theory.  A naked singularity that cannot be so hidden may indicate a
pathology---the absence of a well-defined dual field theory, or a
field theory in an unphysical vacuum state.  If we approach a ``good''
naked singularity by taking a limit of regular black holes, the dual
picture is that the infrared cutoff (finite temperature) is being
removed.  The calculational power of AdS/CFT may be limited in such a
circumstance because of the large curvatures near the singularity.

The second motivation for \Conjecture\ is an examination of states on
the Coulomb branch of ${\cal N}=4$ super-Yang-Mills theory.  Finite
temperature draws Coulomb branch states to the origin of moduli space,
where all operators have vanishing vacuum expectation values.  Thus if
one can demonstrate \Conjecture\ for Coulomb branch states, it is a
check of the conjecture which is orthogonal to finite temperature
considerations.  In fact there is a class of Coulomb branch vacua
which can be described in $d=5$ ${\cal N}=8$ gauged supergravity, and
most of them have naked singularities.  These naked singularities can
be regarded as artifacts of the Kaluza-Klein reduction from ten to
five dimensions provided the ten-dimensional geometry avoids the
aforementioned pathology of D3-branes with negative tension and
negative charge.  As we will show in section~\ref{Coulomb}, the
pathological cases for Coulomb branch flows are precisely those where
\Conjecture\ is violated.

\subsection{Finite Temperature}
\label{FiniteT}

The finite temperature generalization of the ansatz \AnsatzOne\ is 
  \eqn{AnsatzTwo}{\eqalign{
   ds^2 &= e^{2 A(r)} (-h(r) \, dt^2 + d\vec{x}^2) + {dr^2 \over h(r)}  \cr
   \vphi &= \vphi(r) \,.
  }}
 Plugging \AnsatzTwo\ into the equations of motion following from \WAction,
we obtain the following:
  \eqn{EOMS}{\eqalign{
   \square\vphi &= h (\vphi'' + 4 A'\vphi') + h' \vphi' = 
    {\partial V \over \partial\vphi}  \cr
   G^t{}_t - G^x{}_x &= -{1 \over 2} (h'' + 4 A' h') = 0  \cr
   G^t{}_t - G^r{}_r &= 3hA'' = -2 h\vphi'^2 
  }}
 and the ``zero energy'' constraint
  \eqn{ZeroEnergy}{
   G^r{}_r = {3 \over 2} A' (h' + 4 A' h) = h \vphi'^2 - 
    2 V(\vphi) \,.
  }
 Differentiating \ZeroEnergy\ with respect to $r$ yields an equation which
follows algebraically from \EOMS, as $r$-reparametrization invariance
demands.

We can solve for $A(r)$ and $h(r)$ in terms of $\vphi'$ using only
quadratures:
  \eqn{Quadratures}{\eqalign{
   A(r) &= {r \over L} - \int_r^\infty dr_1 \int_{r_1}^\infty dr_2 \,
    {2 \over 3} \vphi'(r_2)^2  \cr
   h(r) &= 1 - \int_r^\infty dr_1 \, {4B \over L} e^{-4 A(r_1)} \,.
  }}
 In \Quadratures\ we have specified some boundary conditions at $r \to
\infty$.  Choosing $\lim_{r\to\infty} h(r)=1$ near the boundary removes the
freedom to rescale $r$ multiplicatively.  The zero energy constraint fixes
$\lim_{r\to\infty} \left( h(r) A'(r)^2 + {1 \over 3} V(\vphi(r)) \right) =
0$, so if one defines $L$ through the equation $\lim_{r\to\infty}
V(\vphi(r)) = -3/L^2$, the normalization of the linear term in $A(r)$ in
\Quadratures\ is fixed.  We are free to add a constant to $A(r)$, or to
shift $r$ itself by a constant.  One of these freedoms can be absorbed by
making a uniform dilation on $t$ and $\vec{x}$; the other amounts to a
conformal transformation which scales all dimensionful quantities uniformly
according to their dimensions.  The only free integration constant in
\Quadratures\ with physical meaning is $B$, which is essentially a
non-extremality parameter.  If $B$ is negative, $h(r)$ would never be zero,
and we would get solutions which are variants of negative mass
AdS-Schwarzschild.  These we regard as manifestly unphysical.

Suppose now that $B>0$.  The function $h(r)$ can have at most one
zero, because it is monotonic.  If a zero exists, it indicates the
presence of a black hole horizon.  Its location, $r = r_H$, must
satisfy $r_H \geq {L \over 4} \log B$, with equality if and only if no
scalars are excited (that is, for AdS-Schwarzschild only).  The
Hawking temperature and Bekenstein-Hawking entropy (per unit volume in
the $\vec{x}$-directions) are
  \eqn{HawkingT}{\eqalign{
   T &= {e^{A(r_H)} h'(r_H) \over 4\pi} = {B \over \pi L} e^{-3 A(r_H)}  \cr
   {S \over V} &= \pi e^{3 A(r_H)} \,.
  }}
 In a Euclidean version of \AnsatzTwo, with Euclidean time $\tau = it$, the
good coordinates at $r=r_H$ are
  \eqn{GoodCoords}{
   y = 2 \sqrt{r-r_H \over h'(r_H)} \qquad 
   \theta = e^{A(r_H)} {h'(r_H) \over 2} \tau \,.
  }
 In terms of these variables, the $\tau$--$r$ part of the metric near
$r=r_H$ is just $dy^2 + y^2 d\theta^2$.  Taking the $h \to 0$ limit of
the scalar equation of motion, one arrives at the following horizon
boundary conditions $r=r_H$:
  \eqn{HorizonBC}{
   h' \vphi' = {\partial V \over \partial\vphi} \,.
  }
 The other equations of motion do not lead to any further boundary
conditions in the sense of specifying a well-defined boundary value
problem.  The condition \HorizonBC\ is actually more informative than
the obvious requirement that $\vphi$ has to be smooth in the good
variables \GoodCoords.  In those variables, \HorizonBC\ becomes ${2
\over y} {d\vphi \over dy} = {\partial V \over \partial\vphi}$.

Existence of solutions of the form \AnsatzTwo\ with regular black hole
horizons does not contradict no-hair theorems.  The reason is that the
scalars $\vphi$ are by assumption tachyons in the asymptotically $AdS_5$
region.  For a particular scalar (call it $\varphi$) the linearized wave
equation near the boundary of $AdS_5$ has two solutions (call them
$\varphi_{(1)}$ and $\varphi_{(2)}$).  Both of them decay as one approaches
the boundary.  The black hole horizon boundary conditions, \HorizonBC, lead
to $n$ constraints if there are $n$ scalars.  But there are $2n$
integration constants for the scalars, leaving us with the expectation of
an $n$ parameter family of solutions.

In this section we will obtain necessary conditions for the existence
of black hole solutions of the form \AnsatzTwo.  In particular, given
a Poincar\'e invariant solution which violates \Conjecture, we will
show that there does {\it not} exist a family of black hole solutions
which can be made to approximate the Poincar\'e invariant solution
arbitrarily closely.  This amounts to the statement that the Poinca\'e
invariant solution does not have near-extremal generalizations.  In
fact, more stringent conditions than \Conjecture\ are probably
necessary in order for a singular geometry to have near-extremal
generalizations.  Conjectures regarding what form those conditions
might take will be presented in sections~\ref{Asymptotics}
and~\ref{Discussion}.

Suppose one starts with a Poincar\'e invariant, singular solution to
\FirstOrder, which can be proven not to have near-extremal
generalizations in five-dimensional classical supergravity.  We wish
to conclude that the dual field theory cannot support finite
temperature.  This is a tricky claim to make, because the curvature
singularity interferes with our ability to compute reliably in
classical supergravity.  Two points in favor of it are as follows.  1)
Assuming there is a ``microscopic'' description of finite temperature
at a singularity which applies beyond the applicability of
supergravity, there should be a smooth match from the microscopic
description to the black hole horizon description at some crossover
point \cite{hpOne}.  2) Numerical studies of two examples for which
the field theory dual is reasonably well-understood indeed exhibit a
smooth limit in which black hole solutions converge to a Poincar\'e
invariant solution.  We will briefly report on the numerical work in
section~\ref{Discussion}, where there is also a further discussion of
the potential pitfalls of identifying solutions that can support
finite temperature with those which admit near-extremal
generalizations.  For now we will content ourselves with an analysis
of what the existence of a static black hole horizon implies.

Let us start with a regular black hole solution of the form
\AnsatzTwo.  For the moment let us not assume that the solution is
asymptotically $AdS_5$.  Evaluating the zero-energy constraint
\ZeroEnergy\ at the horizon ($h=0$), we see that $A' h' = -{4 \over 3}
V(\vphi)$.  This implies $V(\vphi(r_H)) < 0$.  The reason is that both
$A'$ and $h'$ are everywhere positive.  In the case of $A'$, this
follows from the property $A'' \leq 0$ (obvious from the third line of
\EOMS), and the assumption that $A'(r)$ is positive {\it somewhere}
outside the horizon.  In the case of $h'$, it follows once one has
ruled out the solutions akin to negative mass Schwarzschild: $e^{4A}
h'$ is a constant, and to avoid negative mass it has to be a positive
constant.  This derivation scarcely relies on the matter lagrangian:
$A'' \leq 0$ is in fact implied by the null energy condition, just as
in the extremal solution \cite{fgpwOne}.  The conclusion, then, from
the zero energy constraint, is that the bulk cosmological constant
cannot be positive at the black hole horizon.  The only way to
saturate this bound is to take $A'$ and $h''$ identically $0$.  This
gives a rather trivial solution to the vacuum Einstein equations with
zero cosmological constant.

It is in fact possible to improve on the bound $V(\vphi(r_H)) < 0$ if
one is willing to use properties of the matter lagrangian and the
assumption that the spacetime is asymptotically $AdS_5$.  Note that
  \eqn{VDiffDecrease}{\eqalign{
   {d \over dr} \left( V(\vphi(r)) - {1 \over 2} h(r) \vphi'(r)^2 \right) 
      &= {\partial V \over \partial\vphi} \vphi' - {1 \over 2} h'\vphi'^2 - 
         h \vphi' \vphi''  \cr
      &= {1 \over 2} (h' + 8 A' h) \vphi'^2 \geq 0 \qquad 
    \hbox{if $h \geq 0$,}
  }}
 where to obtain the second equality we have used the scalar equation
of motion.  At the black hole horizon, $h=0$ and $\vphi'$ is finite.
At the boundary of $AdS_5$, $h \to 1$ and $\vphi' \to 0$.  So $h
\vphi'^2 = 0$ both at the black hole boundary and at the horizon of
$AdS_5$.  Integrating \VDiffDecrease, we find
  \eqn{VIntDecrease}{
   V(\vphi(r_H)) \leq V_{UV} \,,
  }
 where $V_{UV}$ is the value of $V(\vphi)$ at the ultraviolet fixed
point---usually at $\vphi=0$.  The inequality is saturated only for
AdS-Schwarzschild.  A crucial ingredient in obtaining these
inequalities is that the scalar kinetic term is positive definite.
This is implied by the null energy condition.  The same positivity
properties were the basis of the proofs of the c-theorem in
\cite{gppzOne,fgpwOne}.

In view of \Quadratures, we can specify a solution to \EOMS\ completely
through the pair $(\vphi(r),B)$.  Suppose we have a sequence of regular
black hole solutions, specified as pairs $(\vphi_n(r),B_n)$, with horizon
radii $r_n$, which converges uniformly to a Poincar\'e invariant solution
$\vphi_0(r)$ to \FirstOrder.  More precisely, the convergence criteria are
$B_n \to 0$ and $\sup_{r \geq r_n} |\vphi_n(r)-\vphi_0(r)| \to 0$ as $n \to
\infty$.  Since $B_n \to 0$, the horizon radii $r_n$ will retreat as $n \to
\infty$ to locations with more and more negative $A(r)$---either to the
naked singularity of the Poincar\'e invariant solution, or to $r=-\infty$
if there is no naked singularity.\footnote{A naked singularity has $A(r)
\to -\infty$ at finite $r$.  If $A(r) \to -\infty$ only as $r \to -\infty$,
curvatures could still grow arbitrarily large.  Either way, the geometry
``ends'' when $A \to -\infty$, modulo issues of analytic continuation,
which can be largely disregarded if we think of Wick rotating with respect
to Poincar\'e time rather than some modification of global $AdS_5$ time.}
With minimal assumptions of smoothness on $\vphi_0(r)$ and $V(\vphi)$, we can
deduce from \VIntDecrease\ that
  \eqn{VDecrease}{
   V(\vphi_0(r)) \leq V_{UV}  \qquad \hbox{in the infrared.}
  }
 This is now a constraint on the {\it extremal} solution.

By ``in the infrared,'' we mean that the inequality should hold for
all values of $r$ such that $A(r)$ is sufficiently negative, or at
least asymptotically as $A \to -\infty$.  As before, $V_{UV}$ is the
limit of $V(\vphi(r))$ as we approach the boundary of $AdS_5$.  The
condition \VDecrease\ implies \Conjecture, given that $V$ is a
continuous function of $\vphi$.  It is almost true that \Conjecture\
implies \VDecrease: the simplest way in which one could imagine
\Conjecture\ holding and \VDecrease\ failing is for $\vphi$ to
approach some constant value in the infrared, corresponding to an
infrared fixed point in the dual field theory, such that $V_{IR} >
V_{UV}$---but precisely this possibility is ruled out by the c-theorem
of \cite{gppzOne,fgpwOne}.  We will encounter several examples which
violate both \Conjecture\ and \VDecrease\ by having $V(\vphi(r)) \to
+\infty$ in the infrared.  All of these examples involve naked
singularities, and these particular naked singularities seem very
unphysical.

Before turning to Coulomb branch flows, a remark on different
topologies on the space of field configurations is perhaps in order.
Because $A(r)$ and $h(r)$ can be expressed in terms of $\vphi(r)$ and
$B$ through quadratures, it seems natural to define a norm in terms of
$\vphi$ and $B$ alone.  Given two configurations, $(\vphi,B)$ and
$(\tilde\vphi,\tilde{B})$, a natural Sobolev norm on the difference
$\vphi-\tilde\vphi$ is
  \eqn{Sobolev}{
   \left\Arrowvert \vphi - \tilde\vphi \right\Arrowvert_{H^p_s} = 
    \sum_{k=0}^s \left\Arrowvert \vphi^{(k)} - \tilde\vphi^{(k)} 
     \right\Arrowvert_{L^p} = 
    \sum_{k=0}^s \left( \int_{r_H}^\infty dr \, e^{4 \tilde{A}(r)}
     \left\arrowvert \vphi^{(k)} - \tilde\vphi^{(k)} 
      \right\arrowvert^p \right)^{1/p} 
  }
 for integer $s$.  Here $\vphi^{(k)}$ indicates the $k$-th derivative with
respect to $r$, and we have assumed that the horizon radius $r_H$ for the
configuration $(\vphi,B)$ (defined as the location where $h=0$) is larger
than the horizon radius $\tilde{r}_H$ for $(\tilde\vphi,\tilde{B})$.  We
have chosen the measure $dr \, e^{4 \tilde{A}(r)}$ because it descends
naturally from the five-dimensional volume element $d^5 x \, \sqrt{g}$.
Other measures could be considered if they seem more convenient.  For
$C^\infty$ functions, the $H^\infty_s$ norm is particularly simple:
  \eqn{SobolevInfinity}{
   \left\Arrowvert \vphi - \tilde\vphi \right\Arrowvert_{H^p_s} = 
    \sum_{k=0}^s \sup_{r \geq r_H} \left\arrowvert 
      \vphi^{(k)}(r)-\tilde\vphi^{(k)}(r) \right\arrowvert \,.
  }
 Thus the topology of uniform convergence, which we used to derive
\VDecrease, follows from the $H^\infty_0$ norm, more commonly known as
the $L^\infty$ norm.  $H^\infty_s$ norms are very permissive near the
boundary of $AdS_5$: any relevant deformation of the lagrangian has
finite norm.  The $H_s^2$ norms rule out such deformations for ${\cal
N}=4$ gauge theory, but also they rule out VEV's for dimension~$2$
operators.  It is easy to construct norms which rule out the $r
e^{-2r/L}$ behavior of a dimension~$2$ deformation but permit the
$e^{-2r/L}$ behavior of a dimension~$2$ VEV.  An example is
  \eqn{SuperNorm}{
   \left\Arrowvert \vphi - \tilde\vphi \right\Arrowvert_f = 
    \int_2^\infty dp \, f(p) 
     \left\Arrowvert \vphi - \tilde\vphi \right\Arrowvert_{H_s^p}
  }
 where $f(p)$ is a function which is $1$ near $p=2$ and whose integral over
$(2,\infty)$ is finite.  It would be interesting to see if the topology
induced by such a norm could be used in place of the topology of uniform
convergence for deriving \VDecrease\ as a necessary condition.

\subsection{The Coulomb branch of ${\cal N}=4$ super-Yang-Mills}
\label{Coulomb}

Although it is nice that one can start with an arbitrary black hole
solution and prove definite statements like \VIntDecrease\ that
support the view that the scalar potential must remain bounded above
near a curvature singularity, one may still retain some doubt about
the validity of \Conjecture.  In particular, might it not be that
singularities which violate \Conjecture\ simply correspond to vacua in
the dual field theory which cannot support finite temperature?  In
this section, we will argue against this possibility by considering
explicit examples of such vacua where the dual geometries are well
understood, singular, and in line with \Conjecture.

Next to the maximally supersymmetric $AdS_5$ geometry, the
best-understood solutions to $d=5$ ${\cal N}=8$ gauged supergravity
with $3+1$-dimensional Poincar\'e invariance are those which involve
only the metric and the scalars in the coset $SL(6,{\bf R})/SO(6)$
\cite{klt,fgpwTwo,BrandhuberSfetsosOne,BakasSfetsos,cglp,gStrings}.
The CFT duals (when they are well-defined) are states on the Coulomb
branch of ${\cal N}=4$ super-Yang-Mills which can be parametrized by
the VEV's of operators $\tr X_{(I} X_{J)}$.  These operators,
transforming in the ${\bf 20}'$ of $SO(6)$, are precisely the ones
dual to the $20$ scalars parametrizing $SL(6,{\bf R})/SO(6)$.

There are VEV's for higher dimension operators as well as for those in
the ${\bf 20}'$, and one can calculate them systematically from the
ten-dimensional geometry.  The magic of consistent truncation is that
the profiles of only finitely many scalars capture infinitely many
VEV's.  Roughly speaking, these higher dimension VEV's are sourced by
arbitrary powers of the scalars in the ${\bf 20}'$.  The corresponding
statement in consistent truncation is that the ten-dimensional fields
are determined non-linearly in terms of the five-dimensional fields.

The ten-dimensional geometry is the near-horizon limit of the
background of a continuous distribution of D3-branes:
  \eqn{DThreeBackground}{\eqalign{
   ds^2 &= {1 \over \sqrt{H}} \left( -dt^2 + d\vec{x}^2 \right) +
    \sqrt{H} \sum_{I=1}^6 dy_I^2  \cr
   H &= \int d^6 \xi \, \sigma(\vec\xi) {L^4 \over |\vec{y}-\vec\xi|^4} \,,
  }}
 where the distribution $\sigma$ integrates to $1$.  All the solutions
whose Kaluza-Klein reductions involve only the five-dimensional metric and
the scalars in $SL(6,{\bf R})/SO(6)$ arise from distributions which are
limits of either of the following the following two distributions
\cite{cglp}:
  \eqn{Sigmas}{\eqalign{
   \sigma_6(\vec\xi) &= {1 \over \pi^3\ell_1\cdots\ell_6}
      \delta'\left( 1-\sum_{I=1}^6 {\xi_I^2 \over \ell_I^2} \right)  \cr
   \sigma_5(\vec\xi) &= {1 \over 2\pi^3\ell_1\cdots\ell_5} \Bigg[ -
     \left( 1 - \sum_{I=1}^5{\xi_I^2 \over \ell_I^2} \right)^{-3/2} 
      \Theta\left( 1-\sum_{I=1}^5 {\xi_I^2 \over \ell_I^2} \right)  \cr
    &\qquad{} + 2 \left( 1 - \sum_{I=1}^5 {\xi_I^2 \over \ell_I^2} 
     \right)^{-1/2} \delta\left( 1 - \sum_{I=1}^5 {\xi_I^2 \over \ell_I^2}
      \right) \Bigg] \delta(\xi_6)\,.
  }}
 The distribution $\sigma_5(\vec\xi)$ is the limit of
$\sigma_6(\vec\xi)$ as $\ell_6 \to 0$.  There is a sort of shell
theorem to the effect that shifting all the $\ell_I^2$ by a constant
makes no difference to the geometry away from the branes (see
\cite{klt,ChepelevRoiban,GiddingsRoss} for evidence of this theorem,
and \cite{cglp} for a crisp statement).  Both $\sigma_5$ and
$\sigma_6$ fail to be everywhere positive.  This is the signal that
there are ``ghost'' D3-branes of negative tension and negative charge,
as well as normal D3-branes with positive tension and positive charge.
In the absence of an understanding of the geometries involving
``ghosts'' along the lines of \cite{jpp}, we are inclined to regard
them as unphysical.  If we send $\ell_5 \to 0$ in $\sigma_5$, the
resulting distribution is a delta function shell in the shape of an
ellipsoid:
  \eqn{SigmaFour}{
   \sigma_4(\vec\xi) = {1 \over \pi^2\ell_1\cdots\ell_4} 
    \delta\left( 1 - \sum_{I=1}^4 {\xi_I^2 \over \ell_I^2} \right)
     \delta(\xi_5) \delta(\xi_6) \,.
  }
 This is a positive definite distribution, and so are its limits with one
or more of the remaining $\ell_I$ taken to zero.

Given a distribution $\sigma(\vec\xi)$, vacuum expectation values of
gauge singlet operators can be computed by the rule
  \eqn{SingletVEVs}{
   \langle \tr X_{(I_1} \cdots X_{I_\ell)} \rangle = 
    \int d^6 \xi \, \sigma(\vec\xi) \xi_{I_1} \cdots \xi_{I_\ell}  \,.
  }
 Conversely, a distribution of the form \Sigmas\ (or any limit
thereof) can be reconstructed from the VEV's of the $SO(6)$ irrep
operators $\tr X_{(I_1} \cdots X_{I_\ell)}$, up to an ambiguity which
amounts to shifting all the $\ell_I^2$ by a constant.  Thus the
requirement that the distribution can be made positive definite (by
appropriate choice of this constant shift) must translate into a
series of inequalities among the VEV's for $\tr X_{(I_1} \cdots
X_{I_\ell)}$ which follow from the hermiticity of the individual
$X_I$.  The point of all this is that the ``ghost'' D3-branes are in
this case really unphysical: no local resolution of the geometry which
left the far field form unchanged would change the problematic VEV's.

Let us first take $n$ of the $\ell_I$ equal in \Sigmas, and set the
other $6-n$ to zero.  These cases were studied in detail in
\cite{fgpwTwo}.  The $SO(6)$ global symmetry is broken to $SO(n)
\times SO(6-n)$.  The five-dimensional geometry involves only one
scalar.  More precisely, the trajectory on the scalar manifold is a
geodesic with respect to the sigma-model metric, and we can
parametrize it with a canonically normalized scalar, $\mu$.  The
operator dual to $\mu$ is ${\cal O}_2 = \tr \left[ (6-n) \sum_{I\leq
n} X_I^2 - n \sum_{I>n} X_I^2 \right]$.  Near the boundary of $AdS_5$,
we have $\mu \sim e^{-2r/L}$, corresponding to a positive VEV for
${\cal O}_2$.  Far from the boundary, $\mu$ increases without bound.
For large $\mu$, one finds $W(\mu) \sim -e^{\zeta\mu}$.  The constant
$\zeta$ has values $2/\sqrt{15}$, $\sqrt{2/3}$, $\sqrt{4/3}$,
$\sqrt{8/3}$, and $10/\sqrt{15}$ for $n=1$, $2$, $3$, $4$, and $5$
respectively (see (15) of \cite{fgpwTwo}).  The geometric origin of
these numbers will become clear shortly.

For $n \leq 4$ one finds $V(\mu) \to -\infty$ as $\mu \to \infty$, whereas
for $n=5$ one finds $V(\mu) \to +\infty$.  Thus \Conjecture\ rules out
the pathological $n=5$ case but accepts the physical $n=4$ cases.  This
success, which seems to have nothing to do with finite temperature, is our
second motivation for proposing \Conjecture\ as a good criterion for
distinguishing physical from unphysical geometries.

It is fairly straightforward \cite{BakasSfetsos,cglp} to work out the cases
with unequal $\ell_I$.  Five scalars are involved in general, corresponding
to the five independent eigenvalues of the matrix $M = S S^T$, where $S \in
SL(6,{\bf R})/SO(6)$.  In an appropriate basis, $M = \diag\{
e^{2\beta_1},e^{2\beta_2},\ldots,e^{2\beta_6}\}$, and $\sum_k \beta_k =
0$.  Following \cite{fgpwTwo}, we introduce scalars $\varphi_1$ through
$\varphi_5$ via the relation
  \eqn{PhiRelations}{
      \pmatrix{ \beta_1 \cr \beta_2 \cr \beta_3 \cr \beta_4 \cr \beta_5 \cr 
     \beta_6 } = 
    \pmatrix{ 1/\sqrt{2} & 1/\sqrt{2} & 1/\sqrt{2} & 0 & 1/\sqrt{6} \cr
              1/\sqrt{2} & -1/\sqrt{2} & -1/\sqrt{2} & 0 & 1/\sqrt{6} \cr
              -1/\sqrt{2} & -1/\sqrt{2} & 1/\sqrt{2} & 0 & 1/\sqrt{6} \cr
              -1/\sqrt{2} & 1/\sqrt{2} & -1/\sqrt{2} & 0 & 1/\sqrt{6} \cr
              0 & 0 & 0 & 1 & -\sqrt{2/3} \cr
              0 & 0 & 0 & -1 & -\sqrt{2/3} }
    \pmatrix{ \varphi_1 \cr \varphi_2 \cr \varphi_3 \cr \varphi_4 \cr 
      \varphi_5 } \,.
  }
 The sigma model metric on the five-dimensional space parametrized by
the $\varphi_I$ is $\delta_{IJ}$.  The potential, $V = -{1 \over 8L^2}
\left[ (\tr M)^2 - 2 \tr M^2 \right]$, and the superpotential, $W =
-{1 \over 4} \tr M$, can be written explicitly as sums of
exponentials:
  \eqn{VWSums}{\eqalign{
   V(\vphi) &= \sum_\alpha v_\alpha e^{\vec\eta_\alpha \cdot \vphi}  \cr
   W(\vphi) &= \sum_\alpha w_\alpha e^{\vec\zeta_\alpha \cdot \vphi} \,.
  }}
 All the constants $w_\alpha$ are negative, but the same is not true
of the $v_\alpha$.  The six vectors $\vec\zeta_\alpha$ are of equal
length, and they sum to zero.  The level sets of $W(\vphi)$ for large
$|\vphi|$ are asymptotic to concentric ``hexahedra'' in ${\bf R}^5$.
A hexahedron is the six-sided polyhedron which is the five-dimensional
generalization of the tetrahedron.  It has $\left( 6 \atop n \right)$
$(n-1)$-dimensional ``faces,'' where a zero-dimensional ``face'' is a
vertex, a one-dimensional ``face'' is an edge, and so on up to $n=6$.
The four-dimensional faces of the hexahedra are normal to the
$\vec\zeta_\alpha$.  The geodesic gradient flow trajectories which
preserve $SO(n) \times SO(6-n)$ symmetry are straight lines in ${\bf
R}^5$ which pass through the center of an $n-1$-dimensional face of
each hexahedron; alternatively we may describe them as the rays
emanating from the origin of ${\bf R}^5$ parallel to a sum of $6-n$ of
the vectors $\vec\zeta_\alpha$.  Other gradient trajectories of
$W(\vphi)$, which lift to ten-dimensional geometries with unequal
$\ell_I$, must run asymptotically parallel to one of the geodesics we
have described.  Those which run asymptotically parallel to some
$\vec\zeta_\alpha$ correspond to cases with five non-zero $\ell_I$, and
these are the ones we wish to rule out on account of the pathology of
``ghost'' D3-branes.  This is exactly what the condition \Conjecture\
does.  The region $V(\vphi) < -3/L^2$ includes all the faces of the
hexahedra of dimension less than four, but it excludes more and more
of the four-dimensional faces as $|\vphi|$ increases, so that
asymptotically none of the trajectories perpendicular to these faces
are allowed.  

\begin{figure}[p]
   \centerline{\psfig{figure=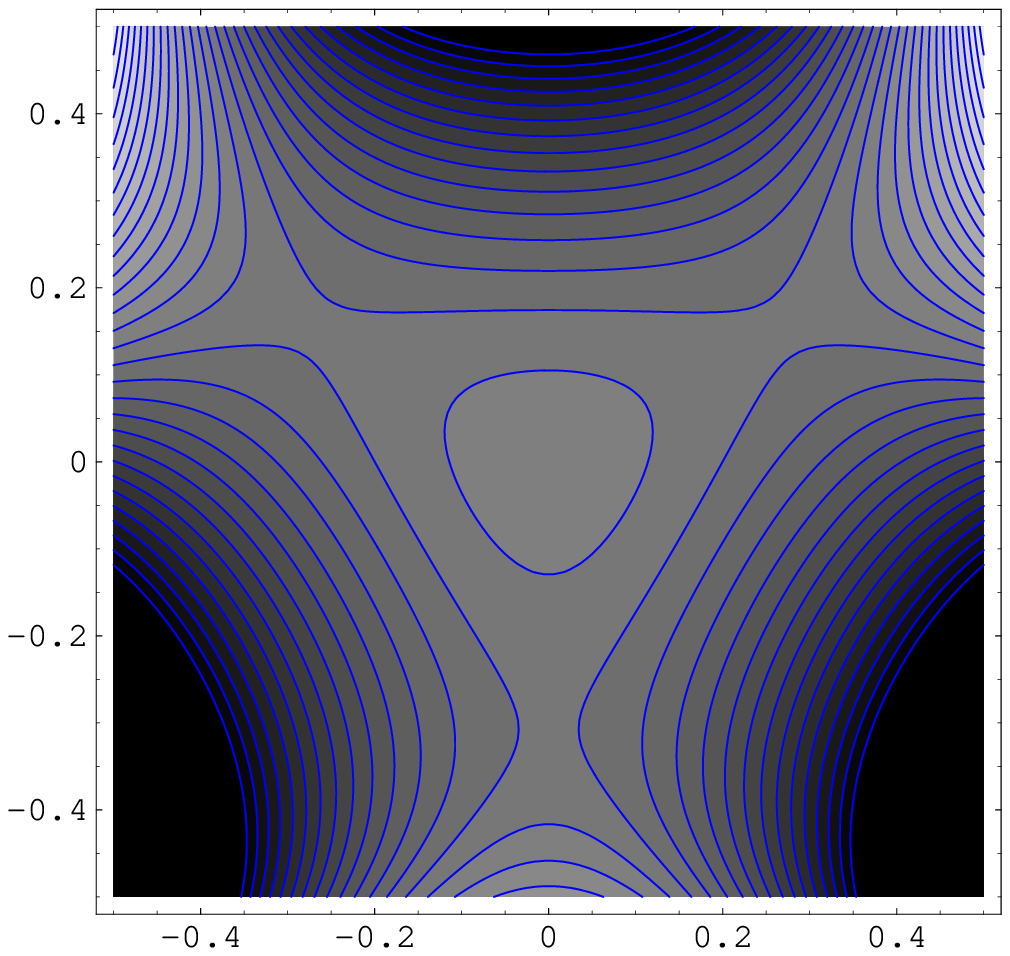,width=2.9in}\qquad
               \psfig{figure=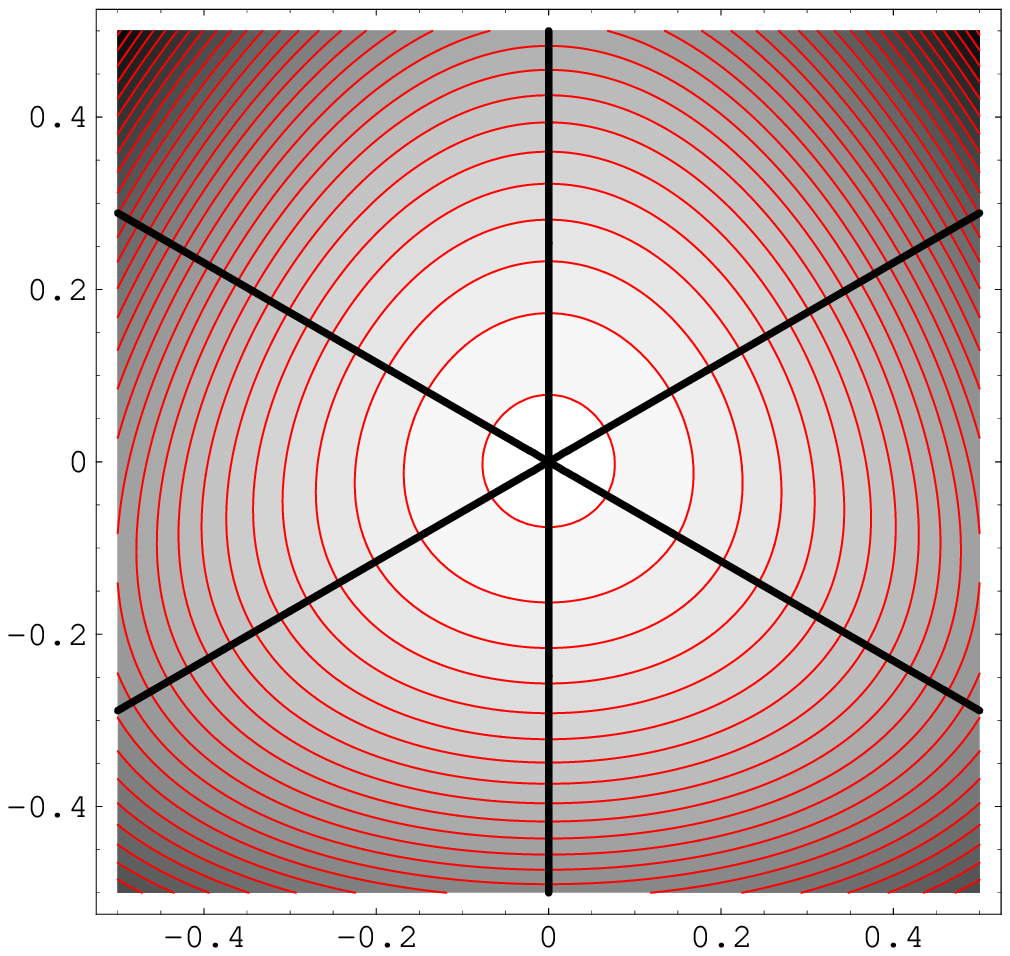,width=2.9in}}
   \centerline{\hskip1in (a) \hskip3in (b) \hskip1in}
   \caption{(a) Contour plot of $V$ and (b) contour plot of $W$, for
$W(\vphi) = -{1 \over 4} e^{\sqrt{3} \varphi_1 + \varphi_2} - {1 \over
4} e^{-\sqrt{3} \varphi_1 + \varphi_2} - {1 \over 4} e^{-2 \varphi_2}$
and $V(\vphi) = {1 \over 8} \left( {\partial W \over \partial\vphi}
\right)^2 - {1 \over 3} W(\vphi)^2$.  $W$ was chosen arbitrarily, but
$V$ and $W$ exhibit the same features in two dimensions that obtain in
five for the $V(\vphi)$ and $W(\vphi)$ that describe Coulomb branch
states for ${\cal N}=4$ super-Yang-Mills.  Regions of higher elevation
are lighter in shade.  If the region of the plots were much larger,
one would see that $V(\vphi) \to +\infty$ as $|\vphi| \to \infty$,
unless one proceeds in one of three special directions separated by
$120^\circ$: north, east-southeast or west-southwest.  The center of
both plots is the local maximum representing the maximally
supersymmetric $AdS_5$ vacuum.  The three saddle points of $V(\vphi)$
are analogs of the unstable $SO(5)$ critical points.  The thick lines
in (b) represent the set ${\cal P}$, to be defined in
\eno{SolutionSet}.  This figure, as well as figures~\ref{figB},
\ref{figC}, and~\ref{figA}, may be easiest to read in
color.}\label{figF}
  \end{figure}

All this may be easier to visualize by extension from
figure~\ref{figF}, which shows a two-dimensional analog with
three-fold symmetry to the five-dimensional example \VWSums\ with
six-fold symmetry.

\subsection{Asymptotics and phase transitions}
\label{Asymptotics}

In all of the examples in section~\ref{Examples}, we will find $V(\vphi)$
and $W(\vphi)$ with the same form as \VWSums, but in some cases some of the
$w_\alpha$ will be positive.  A gradient flow trajectory of $W$ which does
not end at a saddle point will have an asymptotic direction, which we may
specify by a unit vector $\vec{n}$.  Let us parametrize the trajectory by a
canonically normalized scalar $\mu$.  From \FirstOrder\ one can extract a
necessary condition on $\vec{n}$, namely
  \eqn{NecessaryN}{
   -\sum_\alpha w_\alpha \vec\zeta_\alpha e^{\vec\zeta_\alpha \cdot 
     \mu \vec{n}} \, \Big| \Big| \, \vec{n} \qquad \hbox{asymptotically
      as $\mu \to \infty$,}
  }
 where $\vec{u} \, || \, \vec{v}$ means that $\vec{u}$ is parallel to
$\vec{v}$: one is a positive multiple of the other.  There is a
straightforward geometrical construction to identify the solutions to
\NecessaryN.  For each vector $\vec\zeta_\alpha$, define $H_\alpha =
\{ \vphi: \vphi \cdot \vec\zeta_\alpha < 1 \}$.  The intersection $H =
\bigcap_\alpha H_\alpha$ is convex.  It is a compact polyhedron in all
examples we will encounter.  Mathematically, it is similar in
construction to a Brillouin zone, only the vectors $\vec\zeta_\alpha$
do not form a lattice.  Now define a function $\sigma: \partial H \to
\{ -1,0,+1 \}$ as follows.  Given a point $P$ on the boundary
$\partial H$ of $H$, consider all those $H_\alpha$ for which $P$ is
also on the boundary of $H_\alpha$.  If all the associated $w_\alpha$
are negative, let $\sigma(P) = -1$; if all are positive, let
$\sigma(P) = +1$; and if some are positive and some are negative, let
$\sigma(P) = 0$.  Define an apothem of $H$ as a vector with its tail
at the origin and its head on $\partial H$, such that the vector is
orthogonal to the lowest-dimensional face of $H$ which it touches.  We
are again thinking of faces in the generalized sense: they can be
vertices, edges, two-dimensional faces, etc.  The apothems of $H$
include the vectors $\vec\zeta_\alpha/|\vec\zeta_\alpha|^2$, and also
certain linear combinations of these vectors.  If $\vec\kappa$ is an
apothem of $H$ such that $\sigma = -1$ at the head of the apothem,
then there is a solution $\vec{n}$ to \NecessaryN\ which is parallel
to $\vec\kappa$.  And any solution $\vec{n}$ to \NecessaryN\ must be
parallel to an apothem of $H$ for which $\sigma = -1$ or $0$.  Along a
gradient flow trajectory which is asymptotically parallel to an
apothem $\vec\kappa$, we will have $W \sim -e^{\zeta\mu}$ for large
$\mu$, where $\zeta = 1/|\vec\kappa|$.  Since
  \eqn{VWApply}{
   V \approx {1 \over 8} \left( {\partial W \over \partial\mu} \right)^2 - 
    {1 \over 3} W^2 \,,
  }
 we learn that $V \to -\infty$ if $\zeta < \sqrt{8/3}$, and $V \to
+\infty$ if $\zeta > \sqrt{8/3}$.  Thus the condition \Conjecture\
rules out directions in which $|\vec\kappa| < \sqrt{3/8}$.  In the
cases where $\sigma = 0$ at the head of $\vec\kappa$ or where
$|\vec\kappa| = \sqrt{3/8}$, the behaviors of $W$ and $V$ have to be
checked explicitly.

If we imagine $W(\vphi)$ as a mountain, and ourselves as skiers, the
vectors $\vec\kappa$ indicate the possible asymptotic directions that
we could take if we started at $\vphi=0$ and followed the fall line
all the way down.  Vectors with $|\vec\kappa| < \sqrt{3/8}$ indicate
directions which are ``too steep'' for safe skiing and violate
\Conjecture\ as a result.  When all the $w_\alpha$ are negative, the
level curves of $W(\vphi)$ for large $|\vphi|$ approximate concentric
dilated copies of $\partial H$.  When some of the $w_\alpha$ are
positive and some are negative, the level sets are more complicated,
but for large negative or large positive values of $W$ they include
the faces of dilations of $\partial H$ with $\sigma = -1$ or $\sigma =
+1$, respectively.

Just as we constructed the convex polyhedron $H$ and the sign map
$\sigma$ starting with $W$ in the form of a sum of exponentials, we
can construct a convex polyhedron $K$ and a sign map $\rho$ starting
with $V$.  Let $L$ be the union of rays which start at the origin and
pass through a point on $\partial K$ where $\rho = 0$.  The
codimension of $L$ is at least one.  The zero contour of $V$ must be
asymptotic to $L$ for large $|\vphi|$.  It is not guaranteed that
every ray in $L$ is asymptotically close to the zero contour of $V$,
but this is guaranteed for a ray in $L$ associated with a point $P$ on
$\partial K$ such $\rho$ takes both positive and negative values close
to $P$.  The set $L$ divides the space of scalars into wedges which do
or do not violate \Conjecture\ according as $\rho = +1$ or $\rho =
-1$ for the face(s) of $K$ lying in that wedge.  Gradient flow
trajectories which run on or asymptotic to $L$ must be checked
explicitly.

It is transparent from the foregoing discussion that trajectories which
pass \Conjecture\ for potentials of the form \VWSums\ will also satisfy
  \eqn{ProposedCondition}{
   {\partial W \over \partial\vphi} \, \Bigg| \Bigg| \, 
   {\partial V \over \partial\vphi} \qquad \hbox{asymptotically as 
    $A(r) \to -\infty$} \,.
  }
 This condition can also be motivated (although not rigorously) by
finite temperature considerations.  In the far region of a
near-extremal geometry, where $h \approx 1$, the equations \EOMS\ are
approximately solved by a solution to the first order equations,
\FirstOrder.  In the near-horizon region, \HorizonBC\ should apply to
a good approximation.  It is natural to think that the far and near
regions can be matched onto each other provided $\vphi'$ is pointing
roughly in the same direction in both regions.  Modulo some (fairly
strong) technical assumptions, it is possible to show that
\ProposedCondition\ is a necessary condition for the existence of a
sequence of non-extremal solutions which converge to an extremal
solution in the $H^{\infty}_1$ norm.  However, the $H^{\infty}_1$ norm
is a very strong topology in the infrared, and it may be that
near-extremal solutions do not in general converge to an extremal
limit in this norm.

In exploring examples, it is often convenient to look first at a more
restricted category of trajectories than \Conjecture\ allows.  Let us
define
  \eqn{SolutionSet}{
   {\cal P} = \left\{ \vphi: \pm {\partial W \over \partial\vphi} 
     \, \Bigg| \Bigg| \, {\partial V \over \partial\vphi} \right\} \,.
  }
 Given $W$ and $V$ in closed form, it is particularly straightforward
to find ${\cal P}$ because it is the solution set to the equations
$\epsilon^{i_1 \cdots i_n} (\partial W / \partial\varphi^{i_1})
(\partial V / \partial\varphi^{i_2}) = 0$.  Generically, ${\cal P}$
consists of several intersecting curves.  Some of them are asymptotic
to gradient flow trajectories of $W$ in the limit where $A \to
-\infty$.  They can sometimes be excellent approximations to
particular gradient flow trajectories even at finite $A$.  In the
Coulomb branch example of section~\ref{Coulomb}, ${\cal P}$ consists
of lines through the origin passing perpendicularly through every face
(of every dimension) of the hexahedron.  It is thus precisely the
union of all the gradient flow trajectories preserving some $SO(n)
\times SO(6-n)$ symmetry.  Every trajectory runs asymptotically
parallel to some line in ${\cal P}$ in this case.  ${\cal P}$ will not
have quite such nice properties in the examples of
section~\ref{Examples}, but it remains true (as will be seen case by
case) that every trajectory which passes the criterion \Conjecture\ is
also asymptotically parallel to some curve in ${\cal P}$.  It would be
nice to know if there is any deep reason for this.

\begin{figure}
   \centerline{\psfig{figure=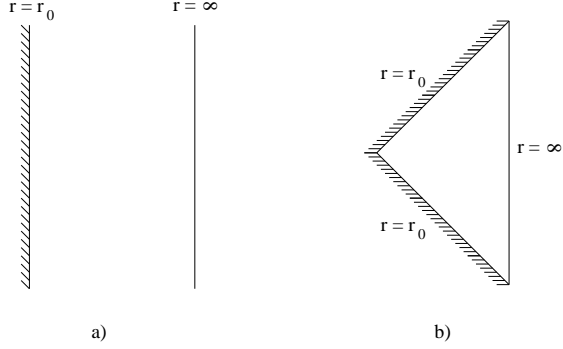,width=2.9in}}
   \caption{(a) The Penrose diagram for $\zeta > \sqrt{2/3}$ is a
strip.  Curvatures diverge as one approaches the right edge.  This is
a naked timelike singularity.  (b) The Penrose diagram for $\zeta \leq
\sqrt{2/3}$ is a wedge.  Curvatures diverge as one approaches either
of the null edges.  This is a null singularity.}\label{figE}
  \end{figure}

It is straightforward to categorize the possible behaviors in the
infrared, provided the scalars' evolution has an asymptotic direction.
As explained above, this assumption is the generic expectation given a
potential $V(\vphi)$ of the form \VWForm.  Let us therefore consider
in more detail the choice $W(\mu) = -e^{\zeta\mu}$.  The variable
$\mu$ is understood as the canonically normalized scalar along the
asymptotic direction of the flow.  Our only aim is to obtain the
correct scaling behavior, so we drop overall factors, including
dimensionful ones like the asymptotic radius of curvature of $AdS_5$.
From \FirstOrder\ one easily obtains
  \eqn{ASoln}{\eqalign{
   ds^2 &= (r-r_0)^{4 \over 3\zeta^2} (-dt^2 + d\vec{x}^2) + dr^2  \cr
   e^{-\zeta\mu(r)} &= {\zeta^2 \over 2} (r-r_0) \,,
  }}
 where $r_0$ is an integration constant which we readily identify as
the location of the singularity.  Curvature invariants involving $n$
derivatives diverge like $(r-r_0)^{-2n}$ as $r \to r_0$.  The causal
structure of the geometry \ASoln\ depends on the value of $\zeta$, and
is more transparent if one introduces a new variable $z$ such that $dr
= -e^A dz$.  Let us assume that the geometry \ASoln\ merges into
asymptotically anti-de Sitter spacetime at large $r$.  Then it is
straightforward to show that if $\zeta > \sqrt{2/3}$ the Penrose
diagram is a strip, as in figure~\ref{figE}(a), while if $\zeta \leq
\sqrt{2/3}$ the Penrose diagram is a wedge, as in
figure~\ref{figE}(b).  Clearly the singularity gets worse as $\zeta$
gets larger: a naked timelike singularity, as in figure~\ref{figE}(a),
is more problematic than a null singularity of the type in
figure~\ref{figE}(b), which is similar to the singular horizon of the
D2-brane.

When $\zeta > \sqrt{8/3}$, $V \to +\infty$ as $r \to r_0$, and the
conjecture \Conjecture\ is that the singularity is unphysical.
Certainly it fails to admit near-extremal generalizations.  When
$\zeta < \sqrt{8/3}$, $V \to -\infty$, and \Conjecture\ is satisfied.
In this case, $|V(\mu)|$ grows slower than $e^{\sqrt{32/3} \mu}$.
Of the solutions $W(\mu)$ to 
  \eqn{VWAgain}{
   {1 \over 8} \left( {\partial W \over \partial\mu} \right)^2 - 
    {1 \over 3} W^2 = V = \left( {\zeta^2 \over 8} - 
     {1 \over 3} \right) e^{2\zeta\mu} \,,
  }
 only $W(\mu) = \pm e^{\zeta\mu}$ grows more slowly than $e^{\sqrt{8/3}
\mu}$ as $\mu \to \infty$.  All the other solutions are asymptotically
of the form $W(\mu) \sim (const) e^{\sqrt{8/3} \mu}$.  It is possible
to integrate \VWAgain\ analytically, but the result is an complicated
implicit equation relating $W$ and $\mu$.  However the asymptotics is
obvious provided $|V(\mu)|$ grows slower than $e^{\sqrt{32/3} \mu}$.
This implies that the generic Poincar\'e-invariant solution to the
equations of motion following from the action~\WAction\ is of the form
\ASoln\ with $\zeta = \sqrt{8/3}$---provided $|V(\vphi)|$ grows slower
than $e^{\sqrt{32/3} |\vphi|}$.  This result has been obtained
previously by a direct analysis of the equations of motion
\cite{gppzThree}.

We have already demonstrated in two different ways that \Conjecture\
is a necessary condition for the existence of near-extremal solutions,
once by using the zero energy constraint plus the null energy
condition, and once by using the scalar equation of motion.  It will
be instructive to present one more proof---inferior to the previous
two because it is less rigorous and less general, but suggestive of
interesting physics at finite temperature.  Start with a Poincar\'e
invariant solution generated using a $W$ which is asymptotic to
$-e^{\zeta\mu}$ with $\zeta > \sqrt{8/3}$.  To obtain near-extremal
generalizations, one would solve \EOMS\ and \ZeroEnergy\ with a small
value of $B$ in \Quadratures.  Let us proceed on the assumption that
for small $B$, the solution for $A(r)$ and $\vphi(r)$ changes only
slightly.  Setting $L=1$ and neglecting factors of order unity, the
expression for $h(r)$ is
  \eqn{hExp}{\eqalign{
   h(r) &= 1 - B \int_r dr_1 \, e^{-4 A(r_1)}  \cr
     &\approx 1 - B \int_r dr_1 \, (r_1-r_0)^{-{8 \over 3\zeta^2}} \,.
  }}
 The integral in the second line remains finite as $r \to r_0$, so it
is indeed true that small $B$ uniformly suppresses the effects of $h$
in the equations of motion.  Thus the approximation used in the second
line of \hExp\ is reliable for sufficiently small $B$.  The conclusion
is that $h(r)$ has no zero for small $B$: there is no horizon!  This
indeed indicates that there are no near-extremal generalizations.  For
sufficiently large $B$, and assuming the criterion \VIntDecrease\ can
be met, it is plausible that solutions with horizons {\it will} exist,
and that they have a temperature which diverges as $B \to \infty$.
Such solutions might be dubbed ``far-extremal,'' since they are not
close to the original Poincar\'e invariant solution.  Assuming the
spacetime is asymptotically $AdS_5$, solutions sufficiently far from
extremality (in the sense of $B$ being large) would be nearly
AdS-Schwarzschild.  Let $B_0$ be the infimum of the set of $B$'s for
which there is a horizon.  It is again plausible that $e^{2A(r)}$ and
$h(r)$ vanish simultaneously when $B=B_0$, and that for $B$ larger
than $B_0$ but sufficiently close to it, $h(r)=0$ at a location
$r=r_H$ where $e^{2A(r)}$ can be made as small as we wish.  Since
$B_0$ is fixed and finite, the formula $T = {B \over \pi L} e^{-3
A(r_H)}$ implies that the temperature diverges as $B \to B_0$.  Thus
there must be some $B_c > B_0$ where $T$ attains its minimum.  (Let us
assume for simplicity that $T$ as a function of $B$ has only one
extremum).  The mass above extremality is related monotonically to
$B$.  Generically one expects that this relation is approximately
linear near $B=B_c$, and that $T \sim T_c + (B-B_c)^2$.  Then as we
approach the critical point from above (both in energy and
temperature), $E-E_c \sim (T-T_c)^{1/2}$, which implies a divergence
in the specific heat, $C \sim (T-T_c)^{-1/2}$.  This is the signal of
a second order phase transition at finite temperature.  On the field
theory side, if one makes the generic assumption that the inverse
square of the correlation length goes to zero like $T-T_c$ (which is
to say, $\nu=1/2$), then the specific heat exponent $\alpha=1/2$
emerges from the hyperscaling relation, $\alpha = 2 - d \nu$ with
$d=3$.  Why the strong interactions should not change the story
drastically is unclear.

If $\zeta < \sqrt{8/3}$, the integral in the second line of \hExp\
diverges as $r \to r_0$, and one might hope to argue on this basis
that near-extremal generalizations of the Poincar\'e invariant
solution do exist.  But when $h$ deviates significantly from $1$, it
is hard to justify the approximation of keeping the same $A(r)$ and
$\vphi(r)$ as in the Poincar\'e invariant solution.  Thus we cannot
make any rigorous claims.  Let us nevertheless attempt to obtain an
equation of state, assuming the asymptotics
  \eqn{AmuForm}{\eqalign{
   A(r) &\sim \alpha_1 \log(r-r_0) + A_0  \cr
   h(r) &\sim 1 - B \int_r dr_1 \, e^{-4 A(r_1)} \,,
  }}
 with $\alpha_1 < 1/4$.  Again we set $L=1$ and neglect factors of
order unity, which is OK since the goal is only to obtain a scaling
relation.  Equation \HawkingT\ may be used to obtain
  \eqn{STForm}{\eqalign{
   B &= e^{4 A_0} (4\alpha_1-1) (r_H-r_0)^{4\alpha_1-1}  \cr
   T &= e^{A_0} (4\alpha_1-1) (r_H-r_0)^{\alpha_1-1}  \cr
   {S \over V} &= e^{3 A_0} (r_H-r_0)^{3\alpha_1} \,.
  }}
 The factors of $e^{A_0}$ indicate the how each quantity scales under
a conformal transformation.  $B$ indeed scales as energy density.
Eliminating $r_H-r_0$ leads to 
  \eqn{STeta}{
   {S \over V} \sim T^\eta  \qquad\hbox{with}\quad
   \eta = {3 \alpha_1 \over \alpha_1 - 1} = 
    {6 \over 2 - 3 \zeta^2}
  }
 where in the last line we have assumed that $\alpha_1 = {2 \over
3\zeta^2}$ as in the Poincar\'e invariant flow, \ASoln.  The
calculation is probably trustworthy when $\zeta$ is sufficiently
small.  Note that as $\zeta \to 0$ one recovers the scaling $S \sim
T^3$ which applies to a conformal field theory.  It is interesting to
note that in the range where $\STeta$ predicts positive specific
heats, it also predicts that $-F/T^4$ is an increasing function of
temperature.  This is consistent with the Appelquist-Cohen-Schmaltz
conjecture \cite{ACS}.\footnote{Despite efforts by the current author
and by R.~Myers, there is no general proof of the monotonicity of
$-F/T^4$ in solutions of the form \AnsatzOne.  I thank R.~Myers for an
extensive correspondence on this topic.}  For $\zeta > \sqrt{2/3}$,
the evidence of \STeta\ is that $T \to \infty$ as $r_H \to r_0$.  It
could be that instead $T$ remains finite in this limit.  Either of
these two outcomes would signal a finite temperature phase transition.
Or it is possible that $T \to 0$ as $r_H \to r_0$, and $S$ as a
function of $T$ falls off faster than any power, indicating a mass
gap.

It would be very interesting to find sufficient conditions for the
existence of black hole solutions arbitrarily close to a given
Poincar\'e invariant flow in an appropriately weak topology.  It would
also be interesting to find more stringent necessary conditions than
\Conjecture.  For instance, one might conjecture that a necessary
condition for a gradient flow to have finite temperature
generalizations which converge to the original geometry in the
topology induced by the norm \SuperNorm\ is that the gradient flow
should either terminate at a saddle point or converge to a curve in
${\cal P}$.  

\begin{figure}
   \centerline{\psfig{figure=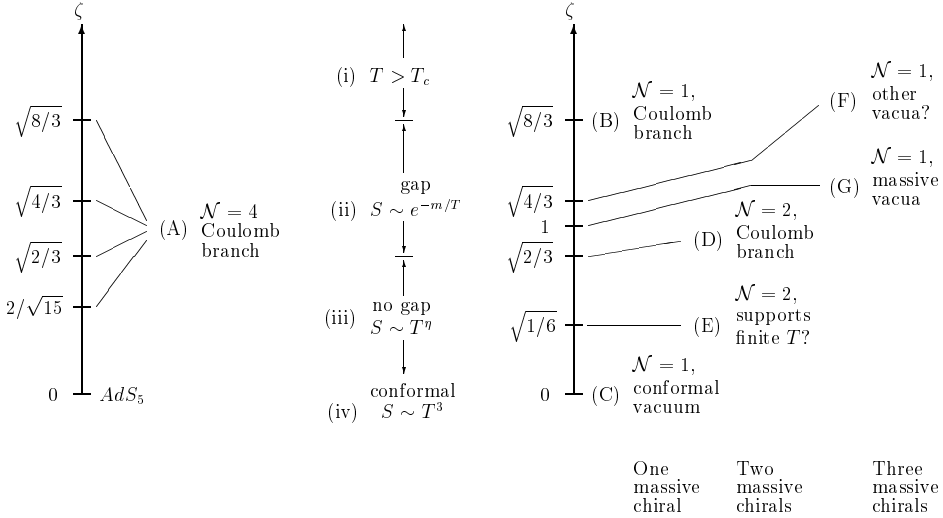,width=4.9in}}
   \caption{Far left: Coulomb branch states exhibit a wide range of
$\zeta$.  Middle: various expectations for the response of
singularities to finite temperature.  Right: a summary of the examples
in section~\ref{Examples}.  Details can be found there regarding each
entry.}\label{figG}
  \end{figure}

Section~\ref{Examples} is concerned with additional examples of
Poincar\'e invariant flows where the field theory physics is more or
less clear.  It is useful to categorize the flows according to the
value of $\zeta$ which controls their far infrared behavior.  This is
done in figure~\ref{figG}.

The responses to finite temperature indicated in figure~\ref{figG} are
based partly on calculation, partly on AdS/CFT examples where the
field theory physics is clear, and partly on conjecture.  The
arguments \hExp\ consitute fairly good evidence that for flows
dominated by some direction in which $\zeta > \sqrt{8/3}$, there is a
finite temperature phase transition.  On the other hand, the
AdS-Schwarzschild solution demonstrates that there are near-extremal
geometries with arbitrarily small temperature when $\zeta=0$.  The
analysis leading to \STeta\ suggests that for $\zeta < \sqrt{2/3}$
there are still near-extremal geometries with arbitrarily small
temperature.  It was found in \cite{fgpwTwo} that minimally coupled
scalars exhibit no gap for geometries of the form \AnsatzOne\ when
$\zeta < \sqrt{2/3}$.  This fits with the power law behavior $S \sim
T^\eta$ found in \STeta.  Note that $\zeta = \sqrt{2/3}$ is also the
value where the Penrose diagram for the Poincar\'e invariant flow
changes from figure~\ref{figE}(b) to figure~\ref{figE}(a): for $\zeta
< \sqrt{2/3}$ the singularity is null, while for $\zeta > \sqrt{2/3}$
it is timelike.  What happens for $\sqrt{2/3} < \zeta < \sqrt{8/3}$ is
harder to understand.  Minimal scalars exhibit a mass gap, so one must
expect that $S$ falls off faster than any power of $T$ for
sufficiently small $T$.  It could be however that curvatures at the
horizon become Planckian before this behavior can be observed.  If so,
the conclusion is simply that a string theory resolution of the
singularity is necessary before the low-temperature physics can be
fully understood.  

The range of $\zeta$'s for the Coulomb branch is puzzling.  How can
this simple system exhibit such a range of behaviors?  One possible
explanation was proposed in \cite{fgpwTwo}.  It is that the
supergravity geometries specify only a continuous distribution of
D3-branes in ten dimensions, and at finite $N$ the physics is most
likely an ensemble average of all discrete distributions which
sufficiently closely approximate the continuous one.  Roughly
speaking, any D3-brane can wiggle as far as its nearest neighbor.
Performing the ensemble average first in the path integral could lead
to extra terms in an effective lagrangian, for instance terms of the
form $(\Sigma^{IJ} \tr X_{(I} X_{J)})^2$ where $\Sigma^{IJ}$ is some
tensor in the ${\bf 20}'$.  These extra effective interactions get
larger as the dimensionality of the brane distribution increases, and
could perhaps explain the peculiar properties that supergravity seems
to ascribe to these configurations.  A naive estimate of the strength
of the double trace terms relative to the original lagrangian is
$N^{1-2/p}$ where $p$ is the dimensionality of the distribution.
Another possibility (also speculative) is that the low-energy theory
of open strings on a $p$-dimensional continuous distribution of
D3-branes is effectively $p+4$ dimensional, simply because the open
strings are allowed to end on any D3-brane.  Then the presence or
absence of a mass gap can be understood in terms of the higher
dimensional theory.\footnote{This interpretation was suggested to me
by J.~Polchinski and J.~Maldacena.}

\section{Examples}
\label{Examples}

The examples of sections~\ref{OneMassive} and~\ref{ThreeMassive} have
appeared previously in the literature \cite{fgpwOne,gppzThree}; the
example in section~\ref{TwoMassive} is based on work with K.~Pilch and
N.~Warner \cite{gpw}.  On the supergravity side, the superpotential
$W$ is associated with supersymmetry transformations: the first order
equations \FirstOrder\ are precisely the conditions for some fraction
of $d=5$ ${\cal N}=8$ supersymmetry to remain unbroken.  As explained
after \VWForm, $W$ is an eigenvalue of a $SO(6) \times SL(2,{\bf
R})$-invariant matrix $W_{ab}$.  Which eigenvalue is the correct one
to describe a given relevant deformation can usually be deduced from
how much supersymmetry is unbroken, plus known asymptotics near the
boundary of $AdS_5$.

On the field theory side, all the examples come from relevant
deformations of ${\cal N}=4$ supersymmetric Yang-Mills theory which
preserve at least ${\cal N}=1$ supersymmetry.  The on-shell spectrum
of ${\cal N}=4$ super-Yang-Mills is a vector, four chiral fermions,
and six real bosons, all in the adjoint of the gauge group:
  \eqn{NFourSpectrum}{\eqalign{
   & \qquad\qquad A_\mu  \cr
   & \qquad \lambda_1\ \lambda_2\ \lambda_3\ \lambda_4  \cr
   &X_1\ X_2\ X_3\ X_4\ X_5\ X_6  \,.
  }}
 These can be grouped into ${\cal N}=1$ multiplets in various equivalent
ways.  The one we will have in mind is $(A_\mu,\lambda_4)$ for the ${\cal
N}=1$ vector multiplet and $(\lambda_1,X_1,X_2)$, $(\lambda_2,X_3,X_4)$,
and $(\lambda_3,X_5,X_6)$ for the adjoint chiral multiplets---also denoted
by chiral superfields $\Phi_1$, $\Phi_2$, $\Phi_3$, whose complex scalar
components are $\phi_1$, $\phi_2$, $\phi_3$.  The superpotential of the
undeformed ${\cal N}=4$ theory is
  \eqn{NFourSuperW}{
   W = \tr \Phi_1 [\Phi_2,\Phi_3] \,.
  }
 Supergravity scalars will typically be denoted $\varphi_\Delta$,
where $\Delta$ indicates the dimension of the corresponding gauge
singlet operator ${\cal O}_\Delta$.  Units will be chosen such that
$L$, the radius of curvature in the asymptotically $AdS_5$ region,
is~$1$.

\subsection{One massive adjoint chiral}
\label{OneMassive}

Our first example is the category of flows that arise from giving a
mass to a single adjoint chiral superfield in the ${\cal N}=4$
langrangian \cite{fgpwOne}.  The mass deformation is specified by
profiles for two scalars, $\varphi_2$ and $\varphi_3$, dual to a boson
mass term ${\cal O}_2$ of dimension~$2$ and a fermion mass term ${\cal
O}_3$ of dimension~$3$.  Explicitly,
  \eqn{OTwoOThree}{\eqalign{
   {\cal O}_2 &= \tr \left( -X_1^2 - X_2^2 - X_3^2 - X_4^2 + 2 X_5^2 +
     2 X_6^2 \right)  \cr
   {\cal O}_3 &= \tr \left( \lambda_4 \lambda_4 + \phi_1 [\phi_2,\phi_3]
     \right) + h.c.
  }}
 The operator ${\cal O}_2$ in \OTwoOThree\ looks like the wrong mass
deformation: adding a positive multiple of it to the lagrangian will result
in negative mass squared for four of the six real scalars.  What we really
aim to do is add $\int d^2 \theta \, {1 \over 2} m \tr \Phi_3^2 + h.c.$ to
the lagrangian.  The operator ${\cal O}_3$ in \OTwoOThree\ is precisely the
dimension~$3$ term in this deformation.  But the dimension~$2$ term is just
$\tr \left( X_5^2 + X_6^2 \right)$.  What saves the day is that
$\varphi_3^2$ can couple to the $SO(6)$ invariant combination $\tr
\sum_{I=1}^6 X_I^2$.  In a supersymmetric background involving the
$e^{-r}$ part of $\varphi_3$ and the $r e^{-2r}$ part of $\varphi_2$,
the $\varphi_3^2$ coupling to the $SO(6)$ singlet mass term must precisely
cancel out the negative terms in the ${\cal O}_2$ of \OTwoOThree, leaving
only the desired $\tr \left( X_5^2 + X_6^2 \right)$.  Similar quadratic
couplings will be important in sections~\ref{TwoMassive}
and~\ref{ThreeMassive}.

The scalars $\varphi_2$ and $\varphi_3$ are canonically normalized, and the
potential and superpotential read \cite{fgpwOne}
  \eqn{VWfgpw}{\eqalign{
   V(\varphi_2,\varphi_3) &= \left[ -\rho^2 + 
    {\cosh 2\varphi_3 - 3 \over 4 \rho^4} + 
    {\rho^8 \over 8} \left( \cosh 2\varphi_3 - 1 \right) \right]
    (\cosh 2\varphi_3 + 1)  \cr
   W(\varphi_2,\varphi_3) &= {1 \over 2} \cosh 2\varphi_3 
    \left( \rho^4 - {2 \over \rho^2} \right) - 
    \left( {3 \over 2} \rho^4 + {1 \over \rho^2} \right)
  }}
 where we have defined $\rho = e^{\varphi_2/\sqrt{6}}$.  It is
consistent to set all other scalars to $0$: locally $V$ is stationary
perpendicular to the $\varphi_2$-$\varphi_3$ plane.  There is a
one-parameter family of flows emanating from the origin, which is the
maximally symmetric point representing unperturbed ${\cal N}=4$ gauge
theory.  The trajectories have only been found numerically.  Their
asymptotics for large $r$ (the ultraviolet) was described in
\cite{fgpwOne}:
  \eqn{PhiAsymptotics}{
   \varphi_3 \sim a_0 e^{-r} \qquad 
    \varphi_2 \sim {4 \over \sqrt{6}} a_0^2 r 
     e^{-2r} + a_1 e^{-2r} \,.
  }
 Unsurprisingly, the leading $r e^{-2r}$ behavior of $\varphi_2$ is fixed
in terms of the leading $e^{-r}$ behavior of $\varphi_3$: this reflects
the fact that the boson mass term and the fermion mass term are related by
supersymmetry.  One can parametrize the gradient flow trajectories with the
quantity
  \eqn{aTildeDef}{
   \tilde{a}_1 = {a_1 \over a_0^2} + {4 \over \sqrt{6}} \log a_0 \,,
  }
 which is invariant under additive shifts of $r$.  It is a dimensionless
measure of $\langle {\cal O}_2 \rangle$ in units set by the scale of the
mass deformation.  $W$ and $V$ are symmetric under $\varphi_3 \to -\varphi_3$,
and from now on we will simplify our discussion by considering only flows
with $\varphi_3 \geq 0$.

\begin{figure}[p]
   \centerline{\psfig{figure=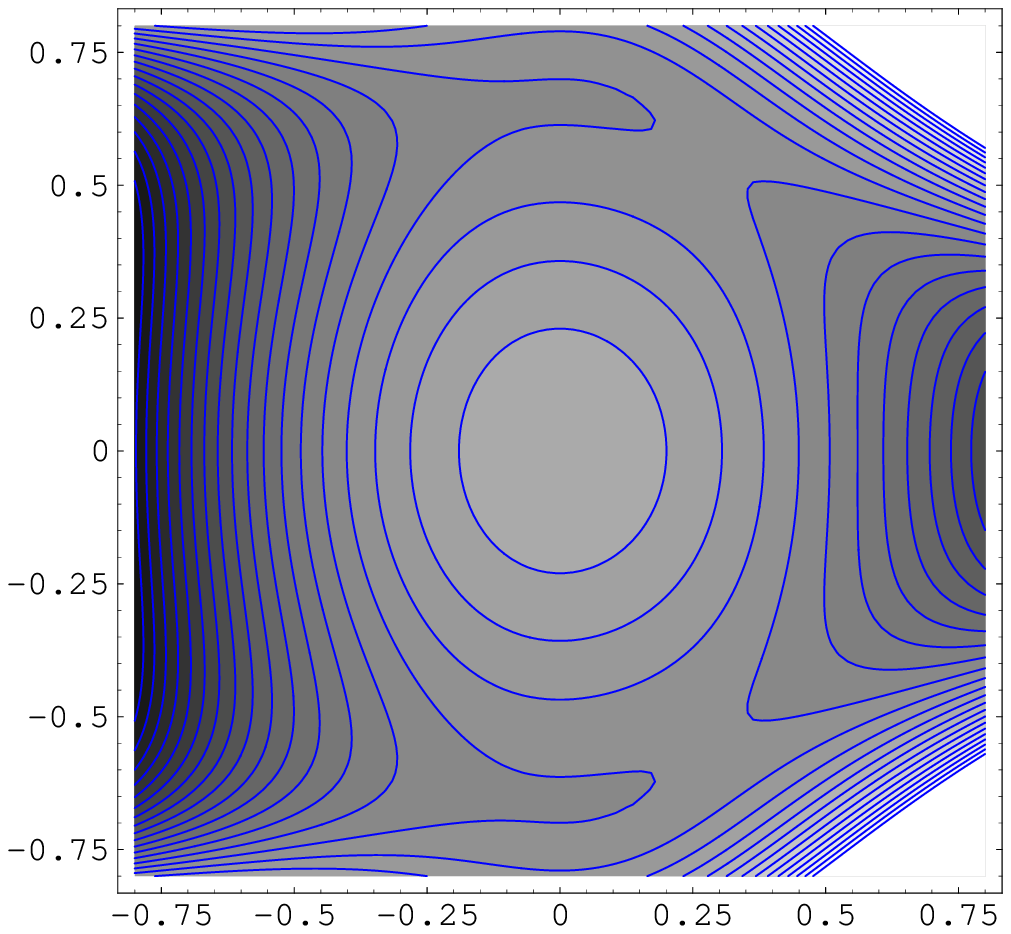,width=2.9in}\qquad
               \psfig{figure=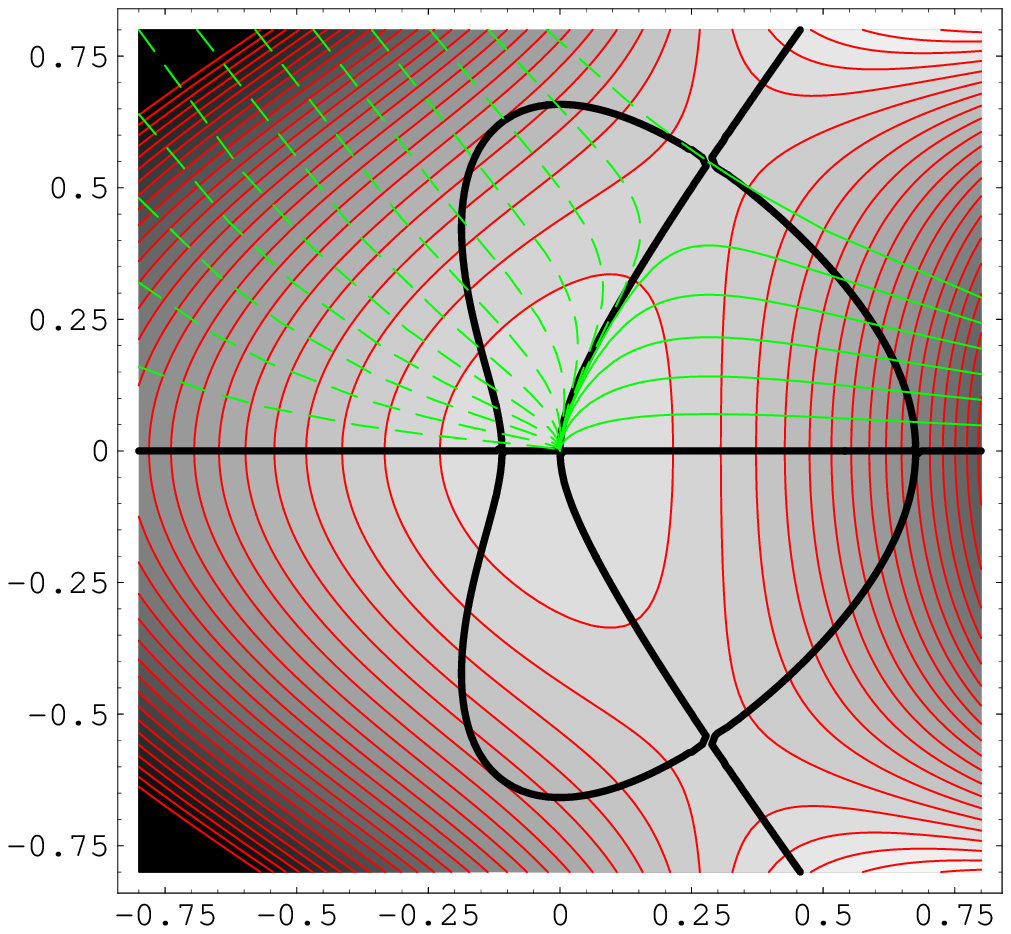,width=2.9in}}
   \centerline{\hskip1in (a) \hskip3in (b) \hskip1in}
   \caption{(a) Contours of $V(\varphi_2,\varphi_3)$, with $\varphi_2$
on the horizontal axis and $\varphi_3$ on the vertical axis.  (b)
Contours of $W(\varphi_2,\varphi_3)$, with the solution set ${\cal P}$
and some typical gradient flow trajectories superimposed.  In both
contour plots, lighter regions correspond to higher elevations.  The
undashed trajectories proceeding asymptotically due east are Coulomb
branch states, and correspond to entry (B) in figure~\ref{figG}.  The
trajectory which ends at the saddle point represents the infrared
conformal fixed point, and corresponds to entry (C) in
figure~\ref{figG}.  The dashed trajectories violate
\Conjecture.}\label{figB}
  \end{figure}

Curiously, it was found in \cite{fgpwOne} that $\tilde{a}_1 =
\tilde{a}^{(c)}_1 = -1.4694\ldots$ corresponds to the conformal vacuum of
the theory, where $\langle {\cal O}_2 \rangle = 0$ and the supergravity
flow terminates at the critical point found in \cite{kpw}.  By comparison
with flows describing states on the Coulomb branch of ${\cal N}=4$ gauge
theory, one can establish that $\tilde{a}_1 > \tilde{a}^{(c)}_1$
corresponds to negative $\langle {\cal O}_2 \rangle$.  Positive $\langle
{\cal O}_2 \rangle$ should be impossible because we have made $X_5$ and
$X_6$ massive.  To put it another way, the infrared fixed point theory has
a Coulomb branch parametrized by color singlet combinations of $X_1$
through $X_4$.  The states with negative $\langle {\cal O}_2 \rangle$ are
$SU(2)_{\rm global} \times U(1)_R$-symmetric states on that Coulomb branch.
States with positive $\langle {\cal O}_2 \rangle$ are unphysical.  In
figure~\ref{figB}(b), the $\langle {\cal O}_2 \rangle < 0$ trajectories are
the ones to the right of the critical trajectory which ends up at the
one-quarter supersymmetric fixed point.  The ones to the left are supposed
to be ruled out.

Felicitously, the condition \Conjecture\ accepts trajectories
with $\tilde{a}_1 \geq \tilde{a}^{(c)}_1$ but rules out those with
$\tilde{a}_1 < \tilde{a}^{(c)}_1$.  The horizontal line along the
$\varphi_2$ axis in the solution set ${\cal P}$ (defined in
\SolutionSet) represents Coulomb branch flows of the ${\cal N}=4$
gauge theory.  When these flows are lifted back to ten dimensions, the
$\varphi_2 > 0$ direction corresponds to an $S^3$ shell of D3-branes
in the hyperplane spanned by $x_1$ through $x_4$, while the $\varphi_2
< 0$ direction corresponds to a uniform disk of D3-branes in the plane
spanned by $x_5$ and $x_6$.  Because the contours of $W$ form a trough
around the $\varphi_2$ axis for large positive $\varphi_2$, the
$\tilde{a}_1 > \tilde{a}^{(c)}_1$ trajectories are drawn to this axis.
The very low energy Coulomb branch physics of these flows does not
depend on whether the adjoints $X_5$ and $X_6$ are given a mass.  In
contrast to this case, the contours of $W$ form a ridge around the
negative $\varphi_2$ axis, and the $\tilde{a}_1 < \tilde{a}^{(c)}_1$
trajectories are repelled from this axis.  These trajectories are
asymptotically parallel to one another, and they are ruled out by
\Conjecture\ because $V(\vphi(r)) \to +\infty$ as $A(r) \to
-\infty$.  The flow exactly along the negative $\varphi_2$ axis
represents sensible ${\cal N}=4$ Coulomb branch physics in the $X_5$
and $X_6$ directions; but this part of the moduli space is lifted
completely when $X_5$ and $X_6$ are given a mass, and the trajectories
are appropriately defocused as a result.

It turns out, as claimed after \SolutionSet, that all the trajectories
permitted by \Conjecture\ run asymptotically parallel to some
curve in ${\cal P}$.  In fact, all trajectories {\it converge} to some
curve in ${\cal P}$ in the far infrared.  The segment of ${\cal P}$
stretching from the origin to the one-quarter supersymmetric saddle
point is an excellent approximation to a gradient flow trajectory.
The part of the curve in ${\cal P}$ which extends northeast from the
one-quarter supersymmetric point does not describe a deformation of
${\cal N}=4$ super-Yang-Mills.

The flows along the positive and negative $\varphi_2$ axes were first
considered in ten dimensions \cite{klt} as zero-temperature, zero-spin
limits of near-extremal D3-brane solutions with some angular momenta
turned on.  Specifically, the positive $\varphi_2$ trajectory comes
from solutions with equal $J_{12}$ and $J_{34}$ turned on, while the
negative $\varphi_2$ trajectory comes from solutions with $J_{56}$.
In five dimensions these angular momenta are interpreted as charges
under the $U(1)^3$ Cartan subalgebra of the $SO(6)$ gauge group, and
the spinning D3-branes are represented as charged black holes
\cite{cgOne,MyersChamblin}.  The time components of the gauge fields
in five-dimensional supergravity act as ``voltages,'' or ``chemical
potentials,'' for conserved $R$-currents in the gauge theory.  The
positive and negative $\varphi_2$ flows are examples of states which
can be approached by tuning these chemical potentials to zero in fixed
ratio with an appropriate power of the temperature.  There is a
thermodynamic instability that sets in at some point in this limiting
process \cite{gSpin}, but that is somewhat outside our current scope
of interest.

\subsection{Two massive adjoint chirals}
\label{TwoMassive}

\begin{figure}[p]
   \centerline{\psfig{figure=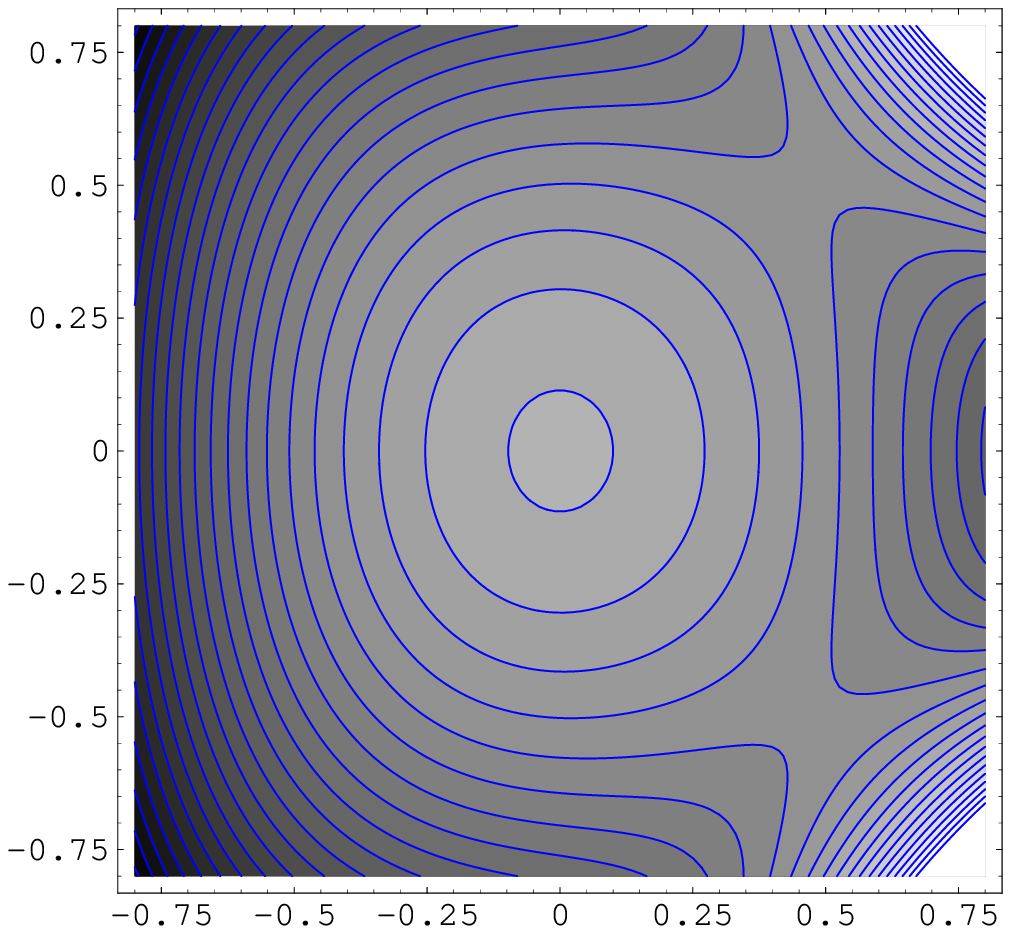,width=2.9in}\qquad
               \psfig{figure=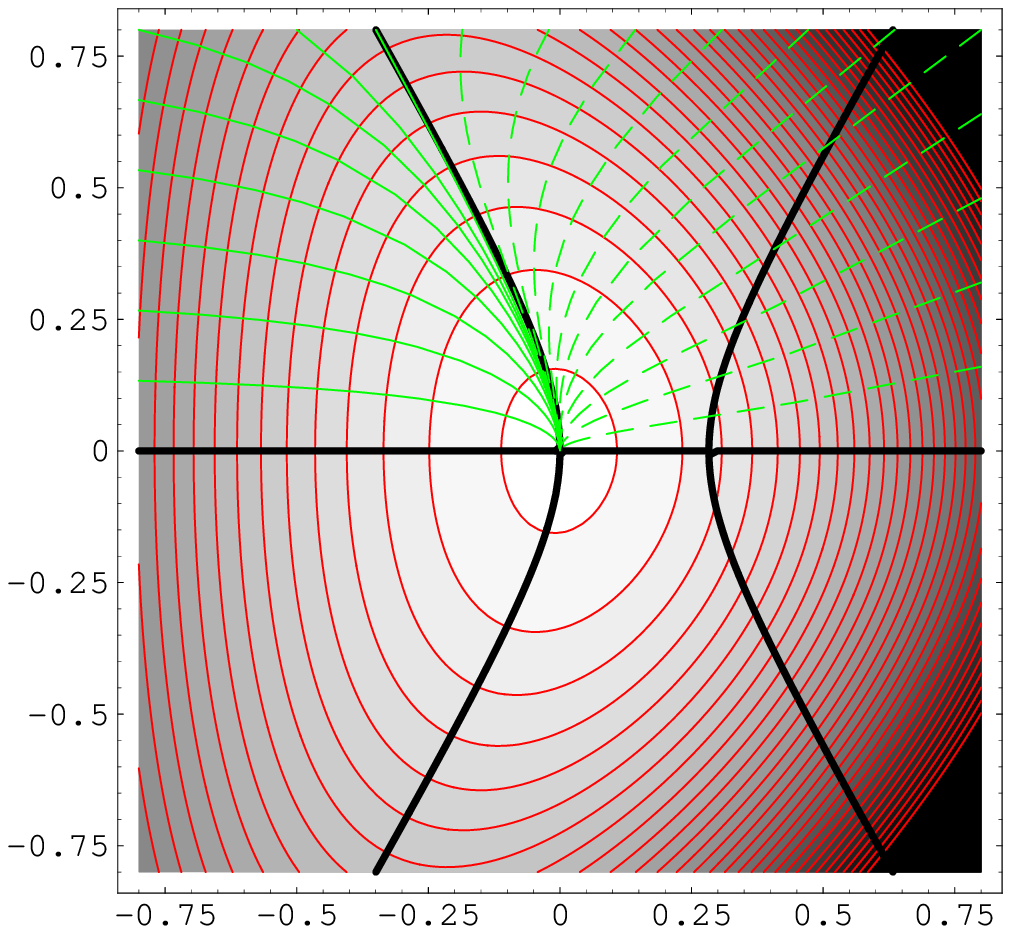,width=2.9in}}
   \centerline{\hskip1in (a) \hskip3in (b) \hskip1in}
   \caption{(a) Contours of $V(\varphi_2,\tilde\varphi_3)$, with
$\varphi_2$ on the horizontal axis and $\tilde\varphi_3$ on the
vertical axis.  (b) Contours of $W(\varphi_2,\varphi_3)$, with the
solution set ${\cal P}$ and some typical gradient flow trajectories
superimposed.  In both contour plots, lighter regions correspond to
higher elevations.  The solid trajectories which proceed
asymptotically due west are states on the Coulomb branch of the ${\cal
N}=2$ gauge theory, and correspond to entry (D) in figure~\ref{figG}.
  The unique solid trajectory which
proceeds northwest nearly along a curve in ${\cal P}$ is the best
candidate for a vacuum which can support finite temperature.  The
dashed trajectories violate \Conjecture.}\label{figC}
  \end{figure}

Our next example is ${\cal N}=4$ super-Yang-Mills broken to ${\cal
N}=2$ by a mass term for an adjoint hypermultiplet.  The analysis is
based on joint work with K.~Pilch and N.~Warner \cite{gpw}.  In ${\cal
N}=1$ language, we are giving equal but opposite masses to two adjoint
chiral multiplets, which we take to be $\Phi_1$ and $\Phi_2$---exactly
the chiral adjoints which we left massless in the previous example.
The field theory was studied in \cite{DonagiWitten}.  The supergravity
\cite{gpw} involves the same scalar $\varphi_2$ in the ${\bf 20}'$ of
$SO(6)$ as we had in the previous example, but a different scalar
$\tilde\varphi_3$ in the ${\bf 10} + \overline{\bf 10}$.  The dual
operators are
  \eqn{OOTilde}{\eqalign{
   {\cal O}_2 &= \tr \left( -X_1^2 - X_2^2 - X_3^2 - X_4^2 + 2 X_5^2 +
     2 X_6^2 \right)  \cr
   \tilde{{\cal O}}_3 &= \tr \left( \lambda_1 \lambda_1 - 
      \lambda_2 \lambda_2 + \hbox{(scalar trilinear)} \right) + h.c.
  }}
 As in the previous example, the square of $\tilde\varphi_3$ can couple to
the $SO(6)$ singlet mass operator.  The combination of this operator and
$-{\cal O}_2$ that supersymmetry demands is proportional to $\tr\left(
X_1^2 + X_2^2 + X_3^2 + X_4^2 \right)$.  Both $\varphi_2$ and
$\tilde\varphi_3$ are canonically normalized, and the potential and
superpotential read
  \eqn{VWTwo}{\eqalign{
   V(\varphi_2,\tilde\varphi_3) &= 
    {1 \over 4} e^{-{4 \over \sqrt{6}} \varphi_2} \left( -4 - 
     8 e^{\sqrt{6} \varphi_2} \cosh \sqrt{2} \tilde\varphi_3 + 
       e^{2 \sqrt{6} \varphi_2} \sinh^2 \sqrt{2} \tilde\varphi_3 \right)  \cr
   W(\varphi_2,\tilde\varphi_3) &=
    -2 e^{-{2 \over \sqrt{6}} \varphi_2} - 
      e^{{4 \over \sqrt{6}} \varphi_2} \cosh \sqrt{2} \tilde\varphi_3 \,.
  }}
 The asymptotics at large $r$ is
  \eqn{LargeR}{
   \tilde\varphi_3 \sim b_0 e^{-r} \qquad
   \varphi_2 \sim -{2 \over \sqrt{6}} b_0^2 r 
    e^{-r} + b_1 e^{-2r} \,,
  }
 and the invariant which parametrizes the one-parameter family of flows
emanating from the origin is
  \eqn{OneParameter}{
   \tilde{b}_1 = {b_1 \over b_0^2} - {2 \over \sqrt{6}} \log b_0 \,.
  }
 In figure~\ref{figC} we exhibit contours of $V$ and $W$, the solution
set ${\cal P}$, and some characteristic gradient flow trajectories
(obtained numerically).  Because of the $\tilde\varphi_3 \to
-\tilde\varphi_3$ symmetry, we can restrict our attention to flows
with $\tilde\varphi_3 \geq 0$.  When $\tilde{b}_1 = \tilde{b}_1^{(c)}
\approx 0.125\ldots$,\footnote{Random jitter near the maximally
supersymmetric point makes this number (and also $\tilde{a}_1^{(c)}$)
surprisingly hard to pin down numerically.}  the flow is asymptotic to
the curve in ${\cal P}$ which follows the obvious northwest ridgeline
of $W$ (see figure~\ref{figC}).  The condition \Conjecture\ allows
this trajectory and all others to the left of it: $\tilde{b}_1 \leq
\tilde{b}_1^{(c)}$.  The trajectories with $\tilde{b}_1 <
\tilde{b}_1^{(c)}$ run asymptotically parallel to the horizontal axis:
this is an explicit example where \ProposedCondition\ is satisfied
without the trajectories actually approaching any curve in ${\cal P}$.
(States on the Coulomb branch which fail to preserve an $SO(n) \times
SO(6-n)$ global symmetry are another example).  The $\tilde{b}_1 =
\tilde{b}_1^{(c)}$ trajectory is well approximated by a curve in
${\cal P}$.  The trajectories with $\tilde{b}_1 > \tilde{b}_1^{(c)}$
proceed down the northeast face of $W$.  As is evident from comparing
figure~\ref{figC}(a) with figure~\ref{figC}(b), this is a case where
\ProposedCondition\ is violated.  No gradient flow trajectory of $W$
ends up at the saddle points near the northeast and southeast corners
of figure~\ref{figG}(a): these points correspond to $AdS_5$ vacua
which break all supersymmetry (see Table~1, entry (iv) of \cite{kpw}).

There is a natural field theory interpretation of these flows,
inspired by \cite{jpp}, though it must be regarded as speculative in
the absence of more detailed tests.  The far infrared theory is
large-$N$ Seiberg-Witten theory, and the Seiberg-Witten curve
\cite{SeibergWitten} is \cite{Argyres,Klemm}
  \eqn{SWCurve}{
   y^2 = \prod_{k=1}^N (x-\xi_k)^2 - \Lambda^{2N} \,,
  }
 where $\Lambda$ is the dynamically generated scale and $\xi_k$ are
parameters determining the position on the moduli space.  In
\cite{jpp} it was suggested that the enhan\c{c}on singularity
corresponds to Seiberg-Witten theory at the ``origin'' of moduli space
$\xi_k = 0$, and that the branch points $x_k = \Lambda e^{i\pi k/N}$
correspond to the positions of branes in a ring around the
enhan\c{c}on.  The $\tilde{b}_1 = \tilde{b}_1^{(c)}$ flow seems most
likely to be dual to this same point in moduli space, where now the
complex variable $x$ parametrizes the directions of the adjoints $X_5$
and $X_6$ which remain massless.  This interpretation is supported by
the fact that the supergravity geometry preserves a $U(1)$ symmetry
which rotates $X_5$ and $X_6$.  Because the $\tilde{b}_1 >
\tilde{b}_1^{(c)}$ flows ``interpolate'' between the ridgeline flow
and the negative $\varphi_2$ axis, and because the negative
$\varphi_2$ axis is known to correspond to a state on the Coulomb
branch of the undeformed ${\cal N}=4$ theory where the D3-branes are
arranged in a uniform disk in the $X_5$--$X_6$ direction, the most
obvious candidate for a field theory dual of the $\tilde{b}_1 >
\tilde{b}_1^{(c)}$ flows is a point on moduli space where the $\xi_i$
are uniformly distributed on a disk.  When the radius of that disk is
much larger than $\Lambda$, the low-energy physics approaches the
Coulomb branch physics of ${\cal N}=4$ gauge theory.

On this interpretation, the trajectories with $\tilde{b}_1 <
\tilde{b}_1^{(c)}$ are unphysical.  Without the mass deformation, the
disk of D3-branes in the $X_5$--$X_6$ directions can shrink to a point
and then re-expand as an $S^3$ shell in the $X_1$ through $X_4$
directions.  This is exactly the picture found in \cite{fgpwTwo}.  But
when we make $X_1$ through $X_4$ massive, this part of the moduli
space disappears, and we are in the position of attempting to give a
negative VEV to $\tr (X_5^2 + X_6^2)$.  All this fits in perfectly
with \Conjecture\ which indeed rules out the $\tilde{b}_1 <
\tilde{b}_1^{(c)}$ trajectories.  To lend further support to the
picture we have presented, it would nice if one could show that finite
temperature in the field theory draws the vacuum state to the point
where all $\xi_k = 0$.  Correspondingly in the supergravity, one would
hope that a horizon with small but finite temperature draws the
scalars onto the $\tilde{b}_1 = \tilde{b}_1^{(c)}$ trajectory.

\subsection{Three massive adjoint chirals}
\label{ThreeMassive}

Our last example is the category of flows believed to be dual to
${\cal N}=4$ super-Yang-Mills deformed by a uniform mass for all three
chiral adjoint superfields \cite{gppzThree}: $W \to W + {m \over 2}
\tr \sum_i \Phi_i^2$.  Naively, the infrared theory is ${\cal N}=1$
super-QCD.  In fact, the model has many vacua, corresponding to
non-commutative VEV's for the massive scalars $\phi_i$ which form a
representation of $SU(2)$ \cite{VafaWitten,DonagiWitten}.

The deformation of the lagrangian consists of a dimension~$3$
operator, which preserves $SO(3)$ and gives a uniform mass to three
adjoint fermions; and a dimension~$2$ operator, which is just the
$SO(6)$ symmetric mass term for all the scalars.  The supergravity
scalar dual to the dimension~$3$ mass term will be denoted
$\varphi_m$.  We will also be interested in the dimension three
operator ${\cal O}_3$ which includes the ${\cal N}=1$ gaugino
bilinear.  The dual scalar will be denoted $\varphi_3$.  This is the
same scalar which was used in section~\ref{OneMassive} to introduce a
mass for one species of fermion.  Both $\varphi_m$ and $\varphi_3$ are
canonically normalized.  The potential and the superpotential are
  \eqn{VWGPPZ}{\eqalign{
   V(\varphi_m,\varphi_3) &= -{3 \over 8} 
    \left( \cosh^2 {2\varphi_m\over \sqrt{3}} + 
    4 \cosh {2\varphi_m\over \sqrt{3}} \cosh 2\varphi_3 - 
     \cosh^2 2\varphi_3 + 4 \right)  \cr
   W(\varphi_m,\varphi_3) &= -{3 \over 2} 
    \left( \cosh {2\varphi_m\over \sqrt{3}} + 
    \cosh 2\varphi_3 \right) \,.
  }}
 A striking feature of this case as compared to the previous two is
that there is only one scalar that takes care of the mass deformation.
In fact this is entirely appropriate: there is no supergravity scalar
which couples linearly to the $SO(6)$-invariant boson mass term, $\tr
\sum_I X_I^2$.  The scalar $\varphi_m$ couples linearly to the
$SO(3)$-invariant fermion mass term, and it can couple quadratically
to the $\tr \sum_I X_I^2$.  Demanding supersymmetry guarantees that
the boson and fermion mass terms will come in the right ratio.

There is a one-parameter family of gradient flow trajectories for the
superpotential in \VWGPPZ, all of which can be displayed in analytic form
\cite{gppzThree}:
  \eqn{AllSolns}{\eqalign{
   A &= {1 \over 2} \log \left[ 2 \sinh (r-C_1) \right] + 
        {1 \over 6} \log \left[ 2 \sinh (3r-C_2) \right]  \cr
   \varphi_m(r) &= {\sqrt{3} \over 2} \log {1 + e^{-r + C_1} \over 
    1 - e^{-r + C_1}}  \cr
   \varphi_3(r) &= {1 \over 2} \log {1 + e^{-3r + C_2} \over
    1 - e^{-3r + C_2}} \,.
  }}
 The behavior $\varphi_m(r) \sim e^{-r}$ for large $r$ indicates
that there is a relevant perturbation of the lagrangian.  The
parameter $C_1$ amounts to setting the scale of this mass deformation.
On the other hand, $\varphi_3(r) \sim e^{-3r}$ signals a VEV for the
operator ${\cal O}_3$.  The invariant quantity which parametrizes the
trajectories is $\lim_{r \to \infty} \varphi_3(r)/\varphi_m(r)^3$.  It is
related to the VEV of ${\cal O}_3$:
  \eqn{VEVOThree}{
   \lim_{r \to \infty} {\varphi_3(r) \over \varphi_m(r)^3} = 
    {1 \over \sqrt{27}} e^{C_2 - 3 C_1} = {c_1 \over N^2}
     {\langle {\cal O}_3 \rangle \over m^3} \,,
  }
 where $\Lambda$ is the dynamical mass scale and $c_1$ is a constant
of order unity which can be determined once the normalization of
${\cal O}_3$ and $\Lambda$ are unambiguously specified.  To understand
the factor of $1/N^2$ in \VEVOThree, recall that there is an overall
factor of $1/G_5$ in front of the classical supergravity action.  When
this action is used to compute a correlator of boundary operators,
three powers of the radius of $AdS_5$ combine with $1/G_5$ to give an
overall factor of $N^2$.  An example is the stress tensor two-point
function: suppressing Lorentz indices, the correlation function that
one computes in the classical supergravity approximation is $\langle
T(x) T(0) \rangle \sim c/x^8$ with $c \sim N^2$.  More generally, the
operators to which canonically normalized supergravity scalars couple
with strength one (that is, without explicit factors of $N$ in the
coupling) have $2$-point functions which scale as $N^2$.  An order~$1$
profile for a supergravity scalar is dual to either an order~$1$
coefficient for the dual operator added to the lagrangian, or an
order~$N^2$ VEV for that operator.  In particular, when $\varphi_3$
and $\varphi_m$ of order $1$, $\langle {\cal O}_3 \rangle$ is of order
$N^2$.  Thus for \VEVOThree\ to be consistent with $c_1$ of order $1$
it is necessary to include the explicit $1/N^2$.

\begin{figure}[p]
   \centerline{\psfig{figure=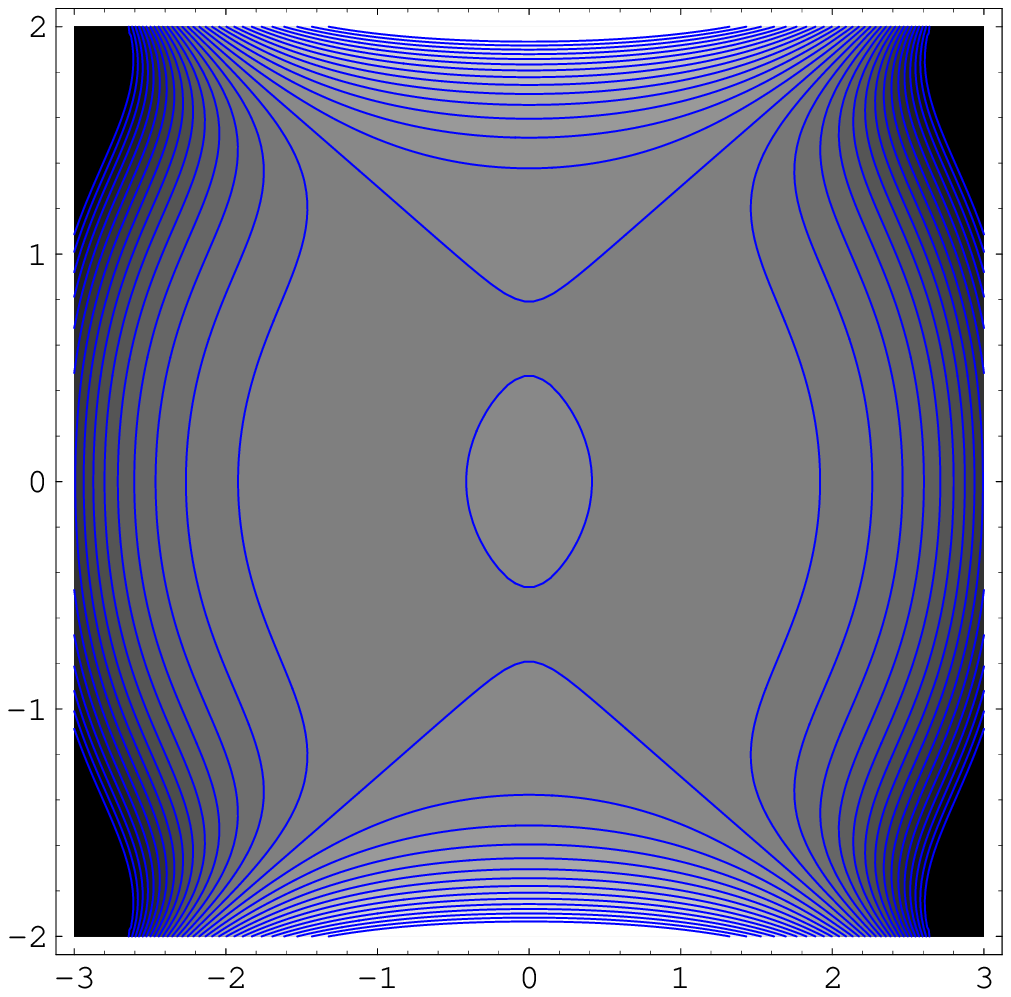,width=2.9in}\qquad
               \psfig{figure=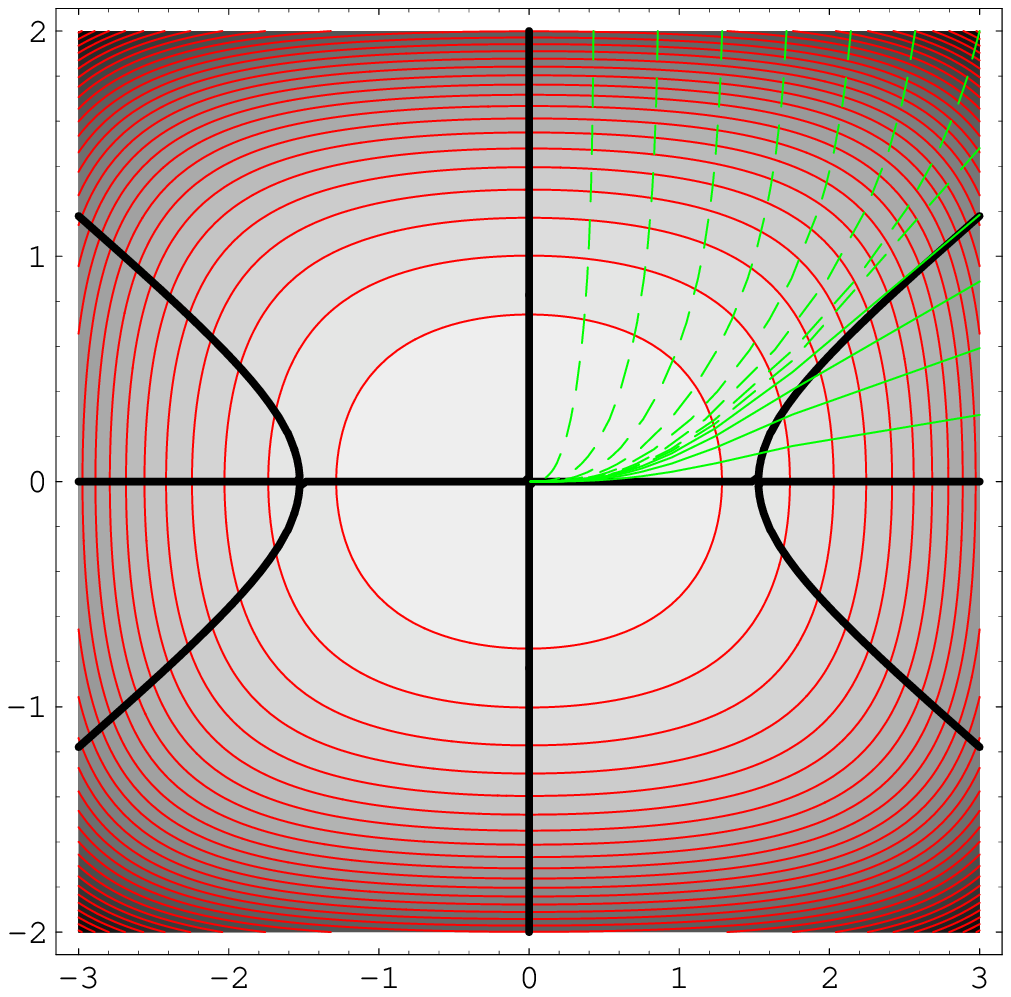,width=2.9in}}
   \centerline{\hskip1in (a) \hskip3in (b) \hskip1in}
 \caption{(a) Contours of $V(\varphi_m,\varphi_3)$, with $\varphi_m$
on the horizontal axis and $\varphi_3$ on the vertical axis.  (b)
Contours of $W(\varphi_m,\varphi_3)$, with the solution set ${\cal P}$
and some typical gradient flow trajectories superimposed.  In both
contour plots, lighter regions correspond to higher elevations.  Note
that the aspect ratio is not unity, so gradient flow trajectories
don't appear to be contour-orthogonal.  The undashed trajectory which
proceeds northeast and is asymptotic to a curve in ${\cal P}$ is the
clearest candidate for a massive vacuum of the theory.  It corresponds
to entry~(G) in figure~\ref{figG}.  The other undashed trajectories
could also be well defined vacua, and correspond to entry~(F) in
figure~\ref{figG}.}\label{figA}
  \end{figure}

On the supergravity side, we can try to use \Conjecture\ to rule out
some trajectories.  Contour plots of $V$ and $W$, the solution set
${\cal P}$, and some typical gradient flow trajectories are shown in
figure~\ref{figA}.
 We can take advantage of the $\varphi_m \to -\varphi_m$ and
$\varphi_3 \to -\varphi_3$ symmetries to restrict our attention to
flows in the northeast corner of the plots.  The
condition~\Conjecture\ allows flows with $C_2 \leq 3 C_1$: the others
violate the steepness condition $\zeta \leq \sqrt{8/3}$.  The curves
in ${\cal P}$ give good approximations to gradient flow trajectories
in the infrared, but the approximation breaks down more badly than in
the previous examples as one proceeds toward the ultraviolet.  The
saddle points on the vertical axis of (a) are the unstable $SU(3)$
critical points.

The field theory is more complicated than in the previous examples,
and it is correspondingly more difficult to identify the vacua
corresponding to different trajectories in figure~\ref{figA}(b).  After
the mass deformation, the superpotential in the field theory is
  \eqn{SuperW}{
   W = \tr \left( {m \over 2} \sum_{i=1}^3 \Phi_i^2 + 
      \Phi_1 [\Phi_2,\Phi_3] \right) \,,
  }
 and the F-flatness conditions read
  \eqn{VaryW}{
   {\partial W \over \partial\phi_k} = m \phi_k + [\phi_{k+1},\phi_{k+2}] 
    = 0 \,,
  }
 where we identify the index $i$ on $\phi_i$ modulo $3$.  Thus the
$\phi_i$ should form some $N$-dimensional representation of $SU(2)$
(not necessarily irreducible).  Imposing the D-flatness conditions and
identifying gauge-equivalent configurations can be achieved
simultaneously by modding out by the action of the complexified gauge
group.  Roughly speaking, one can think of the representations of
$SU(2)$ as non-commutative spheres, whose radii are the quadratic
Casimirs of the representations.  The radius is largest for the
$N$-dimensional irreducible representation.\footnote{Similar
non-commutative configurations were studied in the test-brane
approximation in \cite{MyersDielectric}.  In fact, one can see a hint
of the physics of \VaryW\ by putting D3-branes in a constant
background field $G_{IJK}$, where $G_{(3)}$ is the complex three-form
field strength of type~IIB supergravity, and $IJK$ are indices in the
directions perpendicular to the D3-branes.  $G_{(3)}$ has to be
involved in the ten-dimensional version of the solutions specified in
\AllSolns\ (indeed in the ten-dimensional version of all the solutions
presented in this section with the exception of the Coulomb branch
trajectories of the undeformed ${\cal N}=4$ theory).  If $*_6 \,
G_{(3)} = i G_{(3)}$, where $*_6$ denotes the Hodge dual in the $IJK$
directions, then in the approach of \cite{MyersDielectric} there
remain {\it commutative} flat directions for the D3-branes to spread
out in.  Possibly an improved treatment with non-constant $G_{(3)}$
could approximate the physics of \VaryW\ more closely.}  In this case,
  \eqn{BiggestSphere}{
   \langle \tr \phi_1 [\phi_2,\phi_3] \rangle \approx 
     {N^3 \over 12} m^3 \,.
  }
 Going to the opposite extreme, one could satisfy \VaryW\ trivially by
setting all the $\phi_i$ to $0$.  In this case, the gauge group is
completely unbroken, and there is a gaugino condensate, which
according to the standard analysis is
  \eqn{GauginoCondensate}{
   \langle \tr \lambda_4 \lambda_4 \rangle \approx c_2 e^{2\pi i k/N} 
     N \Lambda^3 \,,
  }
 where $c_2$ is another constant of order unity, $\Lambda$ is the
scale of the low-energy ${\cal N}=1$ theory, and $k$ can run from $1$
to $N$ (see for example
\cite{NSVZ,ShifmanVainshteyn,Konishi,ils,IntriligatorSeiberg,ShifmanTalk}).
Again up to factors of order unity, $\Lambda = m \exp\left( -{8\pi^2
\over 3 g^2 N} \right)$ where $g$ is the gauge coupling at the ${\cal
N}=4$ ultraviolet fixed point.  The overall factor of $N$ in
\GauginoCondensate\ ensures that domain walls between adjacent vacua
have tension scaling as~$N$.  To see this recall that the exact
superpotential has one more power of $N$ than the gaugino condensate;
but domain wall tensions are given by the change of $W$ across the
domain wall, and $W$ changes by a phase $e^{2\pi i/N}$ from one vacuum
to the next \cite{WittenSQCD}.

Naively, the operator ${\cal O}_3$ is $\tr \left( \lambda_4 \lambda_4
+ \phi_1 [\phi_2,\phi_3] \right) + h.c.$, as in \OTwoOThree.  Both the
fermion bilinear and the scalar trilinear are manifestly $SU(3)$
invariant.  The VEV of the first term in the trivial vacuum is
order~$N$, while the VEV of the second is order~$N^3$ in the
irreducible vacuum.  Classical supergravity effects correspond to
VEV's of order $N^2$, so the first behavior appears unobservably
small, while the second seems unobservably large!  We suggest however
that
  \eqn{OThreeAgain}{
   {\cal O}_3 = \tr \left( \lambda_4 \lambda_4 + 
    \sum_{i=1}^3 \phi_i {\partial W \over \partial\phi_i} \right) + 
    h.c. \,,
  }
 where $W$ is the {\it deformed} superpotential, \SuperW, and we are
not attempting to be precise about factors of order unity on either
the fermion or scalar terms.  Classically, the scalar terms have
vanishing VEV in any vacuum.  However, the Konishi relation gives
  \eqn{Konishi}{
   \sum_{i=1}^3 \left\langle
     \tr \phi_i {\partial W \over \partial\phi_i} \right\rangle =
     N \langle \tr \lambda_4 \lambda_4 \rangle \,.
  }
 Thus if $\langle \tr \lambda_4 \lambda_4 \rangle$ is order~$N$, as
the standard field theory analysis indicates, then $\langle {\cal O}_3
\rangle$ is order~$N^2$.

Because the standard field theory analysis of the trivial confining
vacua indicates a VEV for ${\cal O}_3$ which is of the right order to
be observed as a classical effect in supergravity, the natural guess
is that these vacua (or one of them, say the $k=0$ vacuum) should be
identified as the $C_2 = 3 C_1$ trajectory in figure~\ref{figA}(b).
This trajectory is distinguished in that it is the last one permitted
by the condition \Conjecture.  It has a smaller value of $\zeta$ than
all the other trajectories, and it is asymptotic to a curve in ${\cal
P}$, which makes it the most plausible candidate for a flow with
near-extremal generalizations which do not involve deforming the
lagrangian.

The other trajectories permitted by \Conjecture\ could have one of
several interpretations, and without further investigation we cannot
judge which interpretation is correct.  First, they might be in some
way unstable, and so correspond to no well-defined field theory
vacuum.  Second, they might be massless vacua corresponding to
embeddings of $SU(2)$ in $SU(N)$ which leave factors of $U(1)$
unbroken.  Third, they might be massive vacua where $SU(N)$ is broken
down to some $SU(N/p)$ subgroup by a representation of $SU(2)$ which
can be written as $N/p$ blocks of dimension $p \times p$.  A recent
analysis \cite{Dorey} indicates yet one more possibility.  The claim
of \cite{Dorey} is that the standard analysis \GauginoCondensate\ of
the gaugino condensate is dramatically altered at strong coupling, and
that in the large $N$, large $g_{YM}^2 N$ limit the regular $N$-gon of
vacua for the trivial representation of $SU(2)$ degenerates to points
on a line segment.  These points are not evenly distributed; rather,
  \eqn{DoreyCondense}{
   \langle \tr \lambda_4 \lambda_4 \rangle = N m^3/j^2 \qquad
    \hbox{where $j=1,2,3,\ldots$} \,.  
  }
 We have rescaled $m$ to conform with our conventions, set $g_{YM} =
1$, and dropped various factors of order unity.  Since $\langle \tr
\lambda_4 \lambda_4 \rangle$ still scales as $N$, the previous
arguments that $\langle {\cal O}_3 \rangle$ should be observable as a
classical effect in supergravity still hold, and we might still expect
the $C_2 = 3 C_1$ trajectory to correspond to the $j=1$ vacuum (which
is the large $\lambda$ limit of the $k=0$ vacuum in the standard
analysis \GauginoCondensate).  Then some of the $C_2 < 3 C_1$
trajectories might be $j>1$ vacua.

The trajectories with $C_2 > 3 C_1$ violate \Conjecture\ and run
asymptotically parallel to the $\varphi_3$-only trajectory.  This
trajectory seems to be a particularly clear case of an unphysical
geometry: the lagrangian is undeformed, a gaugino condensate has no
reason to form, the flat directions are $[X_I,X_J] = 0$, and yet we
are trying to give a nonzero VEV to ${\cal O}_3 = \tr (\lambda_4
\lambda_4 + \phi_1 [\phi_2,\phi_3])$.  By extension it is plausible to
rule out all the $C_2 > 3 C_1$ trajectories, since they have the same
behavior in the infrared.  Incidentally, the ``universal solution'' of
Horava-Witten theory compactified on a Calabi-Yau manifold
\cite{OvrutStelle} develops the same singularity as the $C_2 > 3 C_1$
flows if the negative tension ``hidden sector'' end-of-the-world brane
is taken to infinite redshift.  The low-energy limit of the hidden
sector gauge theory is pure ${\cal N}=1$ super-Yang-Mills theory.  The
unpleasant singularity could be due to neglect of the gaugino
condensate in the derivation of this solution.

To sum up, what tells in favor of the condition \Conjecture\ in this
example is that the field theory analysis indicates VEV's for $\langle
{\cal O}_3 \rangle$ which are of order $N^2$ and bounded in magnitude.
This qualitative feature is reproduced by \Conjecture\ because it
implies $C_2 \leq 3 C_1$.  It would be interesting to study the
symmetries of the problem more carefully.  In particular, is there a
$U(1)$ symmetry in the supergravity, broken by the choice of
$\varphi_3$, which arises from the large $N$ limit of the discrete
${\bf Z}_{2N}$ symmetry?  According to \cite{Dorey} this symmetry only
pertains in the small $g_{YM}^2 N$ limit.

\section{Fluctuations}
\label{Fluctuations}

Obvious cases where the criterion \Conjecture\ fails to guarantee a
physical interpretation on the field theory side are flows to critical
points of $V$ which violate the Breitenlohner-Freedman bound
\cite{BF,TownsendBF}.  Two well-known examples are flows to the
$SO(5)$ symmetric critical point of $d=5$ ${\cal N}=8$ gauged
supergravity, and flows to the $SU(3)$ critical point
\cite{gppzOne,dzOne,KPunp,dzTwo}.  We would like to have a
generalization of the Breitenlohner-Freedman bound which applies to
cases which are only asymptotically $AdS_5$ in the ultraviolet.  The
obvious one is to demand that the spectral decomposition of the
Lorentzian two-point function of an arbitrary color singlet operator
should involve only timelike momenta.  This statement can be phrased
in the jargon of ``AdS/QCD'' as the absence of tachyonic glueballs.
More generally we can think of it as stability of the field theory
vacuum: a gauge singlet operator ${\cal O}(0)$, acting at the origin
of position space on the vacuum of the dual gauge theory, should
produce only states with timelike momenta.  On Wick rotating to
Euclidean signature, the requirement amounts to having two-point
functions which are non-oscillatory at large distances, but instead
fall off like powers or exponentials.  This makes it clear that the
criterion is indeed a generalization of the Breitenlohner-Freedman
bound, since violations of that bound translate precisely into gauge
singlet operators with complex dimension.

The restriction to timelike momenta is straightforward to implement in
AdS/CFT because the spectrum of momenta-squared is identical to the
spectrum of an appropriately defined one-dimensional Schrodinger
hamiltonian.  This identification is a direct consequence of the
prescription of \cite{gkPol,witHolOne}, and it is the basis of all
AdS-glueball calculations following \cite{witHolTwo,coot}.  Since it
has been discussed, implicitly or explicitly, in many other places in
the literature (for instance
\cite{witHolTwo,coot,fgpwTwo,BrandhuberSfetsosOne,StringsTalk}) we can
afford to be brief.  Our aim is simply to make the discussion of
\cite{fgpwTwo} more systematic.

Consider a Poincar\'e invariant flow solution of the form \AnsatzOne.  It
is convenient to introduce a new radial variable, $z$, such that $dr = e^A
dz$.  The solution assumes the form
  \eqn{NewForm}{\eqalign{
   ds^2 &= e^{2A(z)} \left( -dt^2 + d\vec{x}^2 + dz^2 \right)  \cr
   \vphi &= \vphi(z) \,.
  }}
 Linearizing the equations of motion around this solution is tricky
because linear fluctuations of the graviton couple to linear
fluctuations of the scalars involved in the flow.  However, scalar
fluctuations $\delta\vphi(x^\mu,z)$ satisfying $\delta\vphi(x^\mu,z)
\perp \vphi'(z)$ nearly decouple from the graviton.  That decoupling
would be exact if $\vphi'(z)$ didn't depend on $z$, but we may assume
that $\vphi'$ changes direction only slowly in the deep infrared as
$A(z) \to -\infty$.  Since we are mainly interested in discerning
infrared properties, little is lost in the approximation that the
orthogonal $\delta\vphi(x^\mu,z)$ excitations decouple from the
graviton.  Linearizing the wave equation $\square\vphi = {\partial
V/\partial\vphi}$ leads to
  \eqn{ScalarF}{
   \square\delta\vphi = \left. {\partial^2 V \over \partial\vphi
    \partial\vphi} \right|_{\vphi(z)} \cdot \delta\vphi  \,.
  }
 We can solve via separation of variables: setting $\delta\vphi =
e^{-{3 \over 2} A(z)} e^{ik_\mu x^\mu} \vec\psi(z)$, one obtains
directly
  \eqn{PsiF}{
   \left[ -\partial_z^2 + U(z) \right] \vec\psi = -k^2 \vec\psi \,,
  }
 where
  \eqn{UDef}{
   U(z) = {3 \over 2} A''(z) + {9 \over 4} A'(z)^2 + 
    e^{2 A(z)} \left. {\partial^2 V \over 
    \partial\vphi \partial\vphi} \right|_{\vphi(z)} \,.
  }
 $U(z)$ is a matrix acting in the space of scalars orthogonal to the flow.
The Schrodinger problem \PsiF\ determines the spectrum of $k^2$, and in our
mostly plus conventions the requirement of timelike momenta is $k^2 \leq
0$.  There are two caveats to be kept in mind.  First, we used the
approximation of decoupling from the graviton to derive \PsiF.  Second, if
the flow geometry has a curvature singularity, supergravity gives at best
an approximation to the spectrum.  The curvature singularity indicates an
infrared problem, so in physical cases it should be possible to smooth it
out---for instance by going to finite temperature and Wick rotating so that
the manifold is entirely non-singular.  If the smoothing is done in the far
infrared, it should only affect the behavior of $U(z)$ near the radius
where $A(z) \to -\infty$.  One can hope that both caveats change the
spectrum only in a controllably small way: for instance, they might make
the infimum of the spectrum slightly negative rather than strictly zero.

Let us now check that we recover the Breitenlohner-Freedman bound for AdS
geometries.  Here $e^{A(z)} = L/z$, no scalars are excited, and curvatures
are bounded, so we can impose the sharp inequality
  \eqn{GotBF}{
   \left[ -\partial_z^2 + U(z) \right] = \left[ -\partial_z^2 + 
    {15/4 + m^2 L^2 \over z^2} \right] \geq 0 \,.
  }
 This leads to $m^2 L^2 \geq -4$ by a standard Bessel function analysis.

For geometries with naked singularities, the positivity of the
Schrodinger operator in \PsiF\ is essentially a condition proposed in
\cite{WaldNaked} almost twenty years ago, and also considered in
\cite{HorowitzMarolf}.\footnote{I thank G.~Horowitz for a discussion
of \cite{HorowitzMarolf} and of the basic criterion.}  With trivial
adaptations appropriate to the current context, the contents of
\cite{WaldNaked} can be summarized as follows.  The Schrodinger
operator on the left hand side of \PsiF\ is the radial part of a
linearized wave operator in a static background with a naked
singularity.  Let us use ${\cal A}$ to denote the Schrodinger operator.
${\cal A}$ is symmetric with respect to an appropriate inner product
$(\cdot,\cdot)$ for the wave-functions.  If it is also positive, then
there is a natural extension (the Fredholm extension) of ${\cal A}$ to
a self-adjoint operator on the Hilbert space of functions which is the
closure under the norm $(\vec\psi,\vec\psi) + (\vec\psi,{\cal
A}\vec\psi)$ of smooth functions whose support excludes both the naked
singularity and the boundary of $AdS_5$.  The existence of a
self-adjoint extension of ${\cal A}$ is a natural criterion because it
means that it is possible to define a unitary evolution equation for
linearized fluctuations around the static background.  The crucial
step is positivity of ${\cal A}$, and in AdS/CFT this is equivalent to
the stability of the field theory vacuum.  The bulk spacetime analysis
looks like it could proceed for ${\cal A}$ bounded below (not
necessarily by zero), provided one is willing to introduce a
sufficiently large $(\vec\psi,\vec\psi)$ component in defining the
norm.  But, modulo the caveats mentioned above, the natural AdS/CFT
requirement is positivity.

Suppose there is a naked singularity in the geometry at $z=z_0$.  Let
$u(z)$ denote the smallest (i.e. most negative) eigenvalue of $U(z)$.  Then
through almost the same Bessel function analysis that one uses to translate
\GotBF\ into the Breitenlohner-Freedman bound, one finds that 
  \eqn{FiniteLimInf}{
   \liminf_{z \to z_0} \, (z-z_0)^2 u(z) \geq -{1 \over 4}
  }
 is a necessary condition for ${\cal A}$ to be bounded below.  (This
is almost true; to be rigorous one must exclude the possibility that
$u(z)$ oscillates rapidly).

The conditions we have discussed should generalize to supergravity
fields with spin, and to other dimensions.  It was only to avoid
technical difficulties that we restricted our attention to scalar
fluctuations perpendicular to the flow.  The general problem of
diagonalizing the linearized fluctuations of all supergravity fields
is computationally challenging, but the final requirement is that that
the two-point functions extracted from the diagonalized equations
should involve only timelike momenta in their spectral decomposition.

\section{Discussion}
\label{Discussion}

Applications of AdS/CFT to non-conformal theories appears at present
to be a subject of particulars, with only a few general truths.  The
c-theorem is one such truth \cite{gppzOne,fgpwOne}; the possibility of
replacing second order equations of motion with first order gradient
flow equations is another \cite{fgpwOne,SkenderisTownsend,dfgk,vvOne}.
We have attempted to find a third, namely a criterion for what sorts
of singular behavior are allowed far from the boundary of AdS.  Our
conjectured criterion, \Conjecture, fares reasonably well when
confronted by examples in the literature.  With a few exceptions, it
correctly distinguishes pathological from non-pathological.  The
exceptions are flows to critical points which violate the
Breitenlohner-Freedman bound.  We have suggested that the
Breitenlohner-Freedman bound is a special case of a more general
requirement, namely that the spectral decomposition of two-point
functions in AdS/CFT should involve only timelike momenta.  The
methods proposed for ruling out unphysical singularities---namely,
\Conjecture\ and well-behaved two-point functions---are applicable
independent of supersymmetry.  The examples considered in
section~\ref{Examples} all preserve some fraction of supersymmetry,
but this is only in order to have a clear understanding of the dual
field theory.

The flows to critical points violating the Breitenlohner-Freedman
bound are pathological cases which \Conjecture\ fails to rule out.  In
the current literature, there are no clean exceptions to \Conjecture\
in the other direction---that is, clearly physical solutions which
violate \Conjecture.  It would be very interesting if one could find
such an exception, and to ask how and whether it supports finite
temperature.  In view of the examples in this paper, one's first
question when addressing a putative exception must be, ``Is the theory
in a physical vacuum state?''

Ideally, if the AdS/CFT map were perfectly understood, we would be
able to say precisely which singularities are resolved by non-trivial
infrared physics and which are not.  A complete understanding of this
question might lead to a complicated set of constraints on the
five-dimensional bulk.  The condition \Conjecture\ is a simple
semi-empirical rule that captures some non-trivial aspects of these
constraints.  A particularly clear example is a VEV for the $SU(3)$
singlet operator ${\cal O}_3 = \tr \left( \lambda_4 \lambda_4 + \phi_1
[\phi_2,\phi_3] \right)$.  On the field theory side, such a VEV is
forbidden unless the ${\cal N}=4$ lagrangian is deformed.  The second
term of ${\cal O}_3$ can't have a VEV because the ${\cal N}=4$
effective potential is flat only along directions where the $\phi_i$
commute.  The Konishi anomaly relates the VEV for the first and second
terms of ${\cal O}_3$, so that the vanishing of one implies the
vanishing of the other.  In supergravity, one is certainly free to
perturb $AdS_5$ by a scalar profile indicating a VEV for ${\cal O}_3$,
but at the expense of violating \Conjecture.  The second order bulk
equations have {\it too many} solutions if arbitrary singularities are
allowed; \Conjecture\ is a useful tool for weeding out the good from
the bad.

What precisely is bad about the field theory dual to singularities
that violate \Conjecture\ has to be addressed case by case.  We have
already indicated the problem for $\langle {\cal O}_3 \rangle \neq 0$
without a mass deformation of the ${\cal N}=4$ lagrangian.  In
sections~\ref{OneMassive} and \ref{TwoMassive}, the problem with
trajectories violating \Conjecture\ was that positive definite
operators have negative VEV's.  In section~\ref{Coulomb}, \Conjecture\
ruled out multi-center distributions of D3-branes which involved
``ghosts'' with negative tension and negative charge.  If one formally
uses these ``ghostly'' distributions to compute VEV's of the form
$\langle \tr X_{(I_1} \cdots X_{I_\ell)} \rangle$, the results again
violate inequalities among these VEV's that one can prove using the
hermiticity of the fields $X_I$.

It would be interesting to describe more fully the subset ${\cal S}
\subset E_{6(6)}/USp(8)$ on which the scalar potential is less than
its value at the origin of the coset.  ${\cal S}$ is far from being a
uniform blob or half-space: it has tendrils which come arbitrarily
close to subspaces of positive co-dimension as one proceeds further
and further from the origin of the coset.  We have proven that a
static black hole horizon can only form at a location where the
scalars lie in ${\cal S}$.  The conjecture \Conjecture\ essentially
says that curvature singularities are allowed only if the scalars
remain in ${\cal S}$ as one approaches the
singularity.  (Technically, we have not ruled out situations
where $|\vphi| \to \infty$ and $V(\vphi)$ approaches a constant value
which is larger than $V$ at the origin of the coset.  No example in
the current literature has this feature, and it may be excludable on
general grounds.)

A black hole horizon is the only purely geometrical expression of
finite temperature.  Black holes are also in the generic expression of
finite temperature in a theory with gravity, in the sense that thermal
excitations with enough energy should form a black hole.  It sounds
reasonable to claim that any finite temperature applied to a bulk
geometry with a singularity where $g_{tt} \to 0$ will result in a
black hole horizon cloaking the singularity, simply because the proper
temperature diverges near the singularity.\footnote{This line of
argument was suggested to me by L.~Susskind.}  But it might happen
that the horizon forms in a region very close to the singularity where
curvatures are Planckian.  Then low-energy supergravity will have
limited value; indeed the very notion of horizon might have to be
modified.

In an AdS/CFT context, it is straightforward to estimate when the
five-dimensional supergravity treatment breaks down.  In units where
the radius of curvature of $AdS_5$ is unity, the five-dimensional
Planck length scales as $N^{-2/3}$.  In the case where the dual CFT is
${\cal N}=4$ super-Yang-Mills theory, $N$ is the rank of the gauge
group; more generally, $N^2$ is the central charge of the dual CFT.
Curvature invariants near a singularity of the form \ASoln\ become
Planckian when $r-r_0 \lsim N^{-2/3}$.  What is the entropy of a black
hole horizon which forms just before curvatures become Planckian?  The
horizon area per unit volume at $r=r_H$ is $e^{3 A(r_H)}$.  The
entropy formula is actually more complicated than $S = A/4G_5$ when
higher derivative terms become important
\cite{MyersJacobsonOne,MyersJacobsonTwo,WaldIyer}, but to obtain a
rough-and-ready estimate we will stick with the simple area law.  (For
near-extremal D3-branes with no evolving scalars, the evidence of
AdS/CFT is that all the $\alpha'$ corrections only change the area law
by $4/3$ in the $\alpha'/L^2 \to \infty$ limit \cite{gkPeet,gkArk}.
However a ``phase transition'' at finite $\alpha'/L^2$ cannot be ruled
out \cite{LiPhase}.)  In units where $L=1$, we have $1/G_5 = N^2$, so
the entropy per unit volume is roughly $N^2 e^{3 A(r_H)}$.  Assuming
that near-extremal solutions have nearly the same $A(r_H)$ as their
Poincar\'e invariant limit, one finds that the entropy per unit volume
for a horizon at nearly Planckian curvatures scales as $N^{2-{4 \over
3\zeta^2}}$.  This is to be compared with the expected scaling on the
field theory side.  If the field theory confines, then at low energy
there are $O(1)$ degrees of freedom.  A black hole horizon should have
an $O(1)$ entropy if it purports to describe the theory at a
temperature much smaller than the confinement scale.  This is possible
if $\zeta < \sqrt{2/3}$, but precisely for this range there is no gap
for excitations of a minimal scalar!  We conclude that a
five-dimensional supergravity treatment of near-extremal
generalizations of singularities with $\zeta < \sqrt{2/3}$ should
suffice to describe gapless field theory duals down to some minimum
temperature which scales as an inverse power of $N$, but that the
supergravity approximation is insufficient to describe the confining
phase of a gauge theory.  The original proposal of \cite{witHolTwo}
avoids this problem by arriving at a confining gauge theory as the
low-energy limit of a higher-dimensional gauge theory at finite
temperature.  However one is then left with a description where above
the confinement scale the theory is higher dimensional.

We would like to think that the considerations of this paper will
generalize to a string theoretic setting; that the conjecture
\Conjecture\ will carry over naturally to some more refined
restriction on the nature of highly curved solutions; and that string
theory considerations will provide a detailed ``resolution'' of the
supergravity singularities which are allowed.  In particular,
D3-branes in an $H$-field should resolve the singularities found in
section~\ref{ThreeMassive} (entries (F) and (G) in figure~\ref{figG}).
With the supergravity solutions and the brane resolutions both in
hand, one might make an interesting study of the transition between
the deconfined phase (described by a black hole horizon in the
supergravity solution) and the confined phase (described in terms of
the brane resolution).  It is conceivable that the transition will
involve some dramatic alteration of the supergravity solution which we
are simply not equipped to compute without a full understanding of the
brane resolution.  But we prefer the hopeful view that the low-energy
theory, useful in so many ways as a guide to the qualitative features
of the full string theory, will match more or less smoothly onto the
microscopic description.

Matching a given brane resolution back onto supergravity seems sure to
impose definite boundary conditions on the bulk fields near the
singularity, but the process of finding these resolutions and
extracting their supergravity limits could be tedious.  Black hole
solutions of the form \AnsatzTwo\ provide a quick and dirty route to
establishing the boundary conditions at the singularity without going
to the trouble of finding the brane resolution---provided one is
interested in physics that is robust under finite temperature, and
provided one is willing to grant that the low-energy supergravity
description of finite temperature should match onto a microscopic
description smoothly enough to distinguish good singularities from
bad.

Here is the strategy.  Suppose we have $n$ tachyonic scalars,
corresponding to $n$ relevant operators that one might add to the
microscopic lagrangian.  Fix the more singular behavior of all these
tachyons near the boundary of $AdS_5$: this amounts to fixing the
relevant deformations, but not the VEV's.  Impose finite temperature
in the form of a black hole horizon.  The horizon boundary conditions
\HorizonBC\ amount to $n$ more constraints on the solution.  Assuming
the form \Quadratures\ for $A$ and $h$ (with $B$ left free, to be
determined in terms of the temperature) amounts to using residual
coordinate invariance to fix one boundary condition on $A$ and one on
$h$ as $r \to \infty$.  In total, if $T$ is regarded as fixed, there
are $2n+3$ constraints.  There are $2n+4$ integration constants of
integration in the bulk equations of motion, \EOMS, but the
zero-energy constraint \ZeroEnergy\ fixes one.  So the boundary value
problem is exactly determined: generically there are solutions with a
given temperature $T$, but only discretely many.\footnote{At low
temperatures it may become advantageous to use the non-extremality
parameter $B$ as a control variable instead of temperature, since
there are probably cases where $B$ is a multiple valued function of
$T$ but not vice versa.  $B$ scales as energy density above
extremality.}

As $T$ is lowered, there may be phase transitions, both first and
second order, corresponding to discontinuities or bifurcations in the
space of all finite temperature solutions.  Sorting out such
finite-temperature phase transitions would be a fascinating end in
itself, which we have only delved into superficially with the
prediction that $\alpha=1/2$ at a generic bifurcation.  But let us
suppose we have gotten past all the phase transitions, if necessary by
switching our control parameter from $T$ to $B$ (see the footnote).
As $B \to 0$, the horizon retreats into the singularity, but the
scalars' evolution is still perfectly determined.  In short, black
holes of a given temperature impose just enough boundary conditions to
be unique up to discrete choices, and singular Poincar\'e solutions
which are limits of black holes inherit the same property.

In the case of a single scalar $\mu$, we can offer a definite
conjecture regarding the nature of the condition at the singularity.
Assume that $\mu$ becomes large and positive in solutions with small
$T$ (or, more properly, small $B$), and that $V \sim -e^{\eta\mu}$ as
$\mu \to \infty$ for some $\eta < \sqrt{32/3}$.  Then the Poincar\'e
limit of black hole solutions will have the form \ASoln\ with $\zeta =
\eta/2$.  This is a very restrictive condition: the generic Poincar\'e
invariant solution is of the form \ASoln\ with $\zeta = \sqrt{8/3}$.
Requiring the exceptional $\zeta = \eta/2$ behavior amounts to
imposing one boundary condition at the curvature singularity.  We
conjecture that Poincar\'e invariant solutions of this kind and only
of this kind will have near-extremal generalizations.  This is a much
stronger statement than \Conjecture, and the evidence is slimmer.  It
could be phrased as a weak form of Cosmic Censorship: static nakedly
singular solutions are allowed, but a theory with any thermalizable
excitations will seek out the singularity with the smallest possible
$\zeta$.  The smallest possible $\zeta$ will always be less than
$\sqrt{8/3}$ for potentials of the form \VWForm.

Unfortunately, there is no solution involving only one scalar where we
have a really clean understanding of the physics on the dual field
theory side.  However, we have run numerics to check the ``Weak Cosmic
Censorship'' conjecture of the previous paragraph for two interesting
cases.  The first case is a deformation
  \eqn{NFourDef}{
   {\cal L} \to {\cal L} + {1 \over 2} m^2 \tr \left( 
    -X_1^2 - X_2^2 - X_3^2 - X_4^2 + 2 X_5^2 + 2 X_6^2 \right) \,,
  }
 with $m^2 > 0$.  The corresponding $V$ goes as $-e^{\sqrt{2/3} \mu}$
as $\mu \to \infty$ (see \cite{fgpwTwo}).  Black hole solutions appear
to exist for any $T$, and as $T \to 0$ one does see a scaling region
where the form \ASoln\ indeed appears with $\zeta = \sqrt{1/6}$ to
$0.05\%$ accuracy.  Not unexpectedly, the VEV for ${\cal O}_2 = \tr
\left( -X_1^2 - X_2^2 - X_3^2 - X_4^2 + 2 X_5^2 + 2 X_6^2 \right)$
diverges as $T \to 0$.  The second case is a deformation by a uniform
mass for all three chiral adjoints---the case studied in
section~\ref{ThreeMassive}---but with only the scalar dual to the mass
deformation excited, and not the scalar dual to the gaugino
condensate.  The scalar evolution is along the horizontal axis of
figure~\ref{figA}(b).  One might expect the Poincar\'e invariant
solution to be only metastable, with a perturbative instability toward
forming a gaugino condensate.  In fact the normal modes of linear
fluctuations of the supergravity scalar dual to the condensate do not
include tachyons: the Schrodinger equation \PsiF\ in that case has the
form of supersymmetric quantum mechanics with ${\cal Q} = \partial_z +
3 \cot 2z$, and the supersymmetric ground state corresponds to
shifting to a neighboring trajectory in the plane of
figure~\ref{figA}(b).  Provisionally let us allow this case as a
geometry which satisfies all the constraints we were able to put on
solutions which should have field theory duals, even though the
correct vacuum state on the field theory side remains obscure.  At the
least the solution is a candidate for verifying the conjectures of the
previous paragraph, because $V(\mu) \sim -e^{\sqrt{16/3} \mu}$ and the
Poincar\'e invariant, supersymmetric solution is of the form \ASoln\
with $\zeta = \sqrt{4/3}$.  The numerics is somewhat stiffer in this
case, but small $T$ solutions reveal scaling regions where $\zeta =
\sqrt{4/3}$ can be verified to $3\%$ accuracy.

In light of these numerical results, we feel entitled to a speculation
regarding the marginal case $\zeta = \sqrt{8/3}$.  This is in some
sense the generic case, because if $V \sim -e^{\eta\mu}$ with $\eta <
\sqrt{32/3}$, the generic Poincar\'e invariant solution is of the form
\ASoln\ with $\zeta = \sqrt{8/3}$, whereas if $V \sim e^{\eta\mu}$
with $\eta > \sqrt{32/3}$, the condition \Conjecture\ rules out all
solutions.  In the supersymmetric examples in this paper, we
encountered $\zeta = \sqrt{8/3}$ twice: once in a Coulomb branch state
of ${\cal N}=4$ gauge theory where the D3-branes were arranged in a
perfect $S^3$ shell in the transverse dimensions; and again in a
Coulomb branch state of a ${\cal N}=1$ mass deformation which had
essentially the same $S^3$ shell interpretation.  In both cases, the
field theory dual makes it clear that the vacuum won't support finite
temperature.  If one does turn on finite temperature at the same time
as introducing a term in the lagrangian of the form \NFourDef, then
the trajectory may be nearly the same in the space of scalars, but the
dependence on proper distance will be very different near the
singularity: $\zeta = \sqrt{1/6}$ in the infrared, rather than
$\sqrt{8/3}$.  It is natural to speculate that the generic solutions
with $\zeta = \sqrt{8/3}$ are allowed precisely when they correspond
to exploring flat directions in the dual field theory, and that they
can never support finite temperature.  This unfortunately tells
against some supergravity constructions which have been claimed to
exhibit properties of confinement
\cite{minahan,gDil,ksDilTwo,gppzTwo}, but the case for confinement in
those examples was less than airtight.

It is particularly easy to show that the geometries of
\cite{ksDil,gDil}, involving only the metric and the dilaton, cannot
support a black hole horizon.  The scalar equation of motion plus the
horizon boundary conditions imply that the dilaton is everywhere
constant.  The only static black hole geometry involving only the
metric and the dilaton is AdS-Schwarzschild.  This simple argument is
another clue that singularities with $\zeta = \sqrt{8/3}$ cannot
support finite $T$: they have the same scaling as the solutions of
\cite{ksDil,gDil}.

A feature of AdS/CFT which was essential to the proper interpretations
of the examples in this paper is that a profile for a given scalar
field can indicate either a deformation of the lagrangian or a VEV for
some gauge singlet operator.  This dual role of scalar profiles
complicates any attempt to identify the value of a scalar at a given
radius in $AdS_5$ with a coupling in an effective lagrangian for the
dynamics at the corresponding scale.  We must ask: 1) How do we
disentangle the ``VEV'' part of the scalar profile from the
``deformation'' part?  2) Is such a disentanglement really necessary,
or can the effective lagrangian be defined so as to pick out exactly
the combination of VEV and deformation that AdS/CFT prescribes for its
effective coupling?  The issues might become clearer if one focused on
solutions which are limits of black holes with prescribed behaviors
for the scalar profiles corresponding to deformations of the
microscopic lagrangian.  The nearly unique specification of vacuum
state in these solutions may simplify the interpretation of the
holographic renormalization group proposed in \cite{vvOne}.  For
instance, the identification of beta functions with gradients of
scalars \cite{BehrndtFall,vvOne} seems odd in Coulomb branch states of
${\cal N}=4$ gauge theory, but possibly more natural for mass
deformations in some preferred vacuum.  A true acid test for the
subject would be to reproduce some known quantitative features of the
field theory RG, such as the NSVZ exact beta function.

\section*{Acknowledgements}

 This work would not have been possible without extensive
communications with N.\ Warner, K.~Pilch, and D.~Freedman.  I have
also profited greatly from a number of exchanges with R.~Myers
regarding near-extremal geometries, and with H.~Verlinde regarding the
renormalization group in AdS/CFT.  I also thank T.~Banks, G.~Horowitz,
S.~Kachru, I.~Klebanov, B.~Kol, V.~Periwal, B.~Pioline, J.~Polchinski,
S.~Shenker, E.~Silverstein, M.~Strassler, L.~Susskind, and E.~Witten
for useful discussions, and R.~Myers for comments on an early draft.
This research was supported in part by DOE grant~DE-FG02-91ER40671 and
by the Harvard Society of Fellows.

\newpage

\section*{Appendix}
\label{Appendix}

The purpose of this appendix is to remark on brane-world models in
which there is a curvature singularity parallel to the brane on which
visible sector matter exists, at some finite proper distance from it.
Such constructions were first considered in the context of
Horava-Witten theory in \cite{WittenBound}.  They have also appeared
as a variant of the Randall-Sundrum construction (see for instance
\cite{BrandhuberSfetsosTwo}).  Following in the spirit of
\cite{vvTwo}, it was recently proposed by two Stanford groups
\cite{kaloper,kachru} that such geometries could help solve the
cosmological constant problem.  In this appendix we will consider only
the minimal case where the ``Planck brane'' is at one end of the
spacetime, where $g_{tt}$ is finite, and the singularity is at the
other, where $g_{tt} \to 0$.  In the simplest constructions, where
$g_{tt}$ is monotonic, gravity is in some sense localized on the
Planck brane (hence the name).  At least in most Horava-Witten theory
constructions, the visible sector fields live on the Planck brane.
The singularity then represents (or is resolved by) hidden sector
fields which couple only gravitationally to the visible sector.

A crucial issue, both in the general scheme of \cite{vvTwo} and in the
specific proposal of \cite{kaloper,kachru}, is the boundary conditions
at the singularity.  At the level of classical five-dimensional
gravity, the construction of \cite{kaloper,kachru} is a boundary value
problem, where one boundary is the Planck brane and the other boundary
is a naked singularity.  Similar analyses have appeared in many
places, for example \cite{OvrutStelle,BrandhuberSfetsosTwo,dfgk}.  In
\cite{OvrutStelle,dfgk}, spacetime ends on some orbifold fixed plane
before $g_{tt} \to 0$, and boundary conditions are imposed both there
and on the Planck brane.  By contrast, in \cite{kaloper,kachru}, no
boundary conditions are imposed at the singularity where $g_{tt} \to
0$.  In \cite{OvrutStelle,dfgk} (and other similar treatments) the
boundary value problem is sufficiently determined to fix the
four-dimensional cosmological constant, and there is no reason for it
to be small.  In \cite{kaloper,kachru}, there are fewer boundary
conditions.  The boundary value problem is underdetermined if the
four-dimensional cosmological constant is left unspecified; it becomes
determined (up to an multiplicative shift on the warp factor) if one
specifies the four-dimensional cosmological constant.  In particular,
there do exist solutions with $3+1$-dimensional Poincar\'e invariance.

\begin{figure}
   \centerline{\psfig{figure=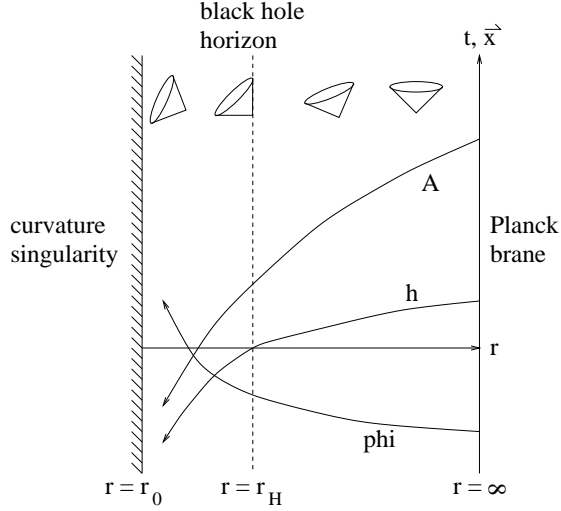,width=2.9in}}
   \caption{The spacetime described by \AnsatzAgain.  This is not a
Penrose diagram: the lightcones tip over as indicated.}\label{figD}
  \end{figure}

The issue which falls within the purview of this paper is whether the
free boundary conditions used in \cite{kaloper,kachru} are reasonable.
Because the solutions there purport to be at least a toy model for
real world physics, they must be able to support finite temperature.
We will interpret this as synonymous with being able to form a black
horizon with finite Hawking temperature close to the singularity,
without drastically changing the rest of the geometry.  (Some
potential pitfalls of this interpretation have been discussed in
section~\ref{Discussion}).  It is straightforward to show that a
horizon is impossible in classical gravity if the bulk scalar is free,
unless in fact the bulk scalar does not vary at all in the solution.
Essentially the relevant observation has already been made in
section~\ref{Discussion}, but we will go into slightly more detail
here to make the case clear.  In order to have finite temperature, we
seek a generalization of the Poincar\'e invariant solution of the form
  \eqn{AnsatzAgain}{\eqalign{
   ds^2 &= e^{2 A(r)} (-h(r) dt^2 + d\vec{x}^2) + dr^2/h(r)  \cr
   \varphi &= \varphi(r)
  }}
 where $h(r)$ is a function which has a simple zero at $r=r_H$, the
black hole horizon.  The temperature is related to $h'(r_H)$ (see
\HawkingT).  A sketch of the geometry is presented in
figure~\ref{figD}.

The scalar equation of motion is (by assumption) $\square\varphi = 0$:
explicitly,
  \eqn{ScalarEOM}{
   e^{-4A} \partial_r e^{4A} h \partial_r \varphi = 0 \,.
  }
 The geometry and the scalars should be perfectly regular at the
horizon: an infalling observer must not notice anything special as he
crosses through it.  Thus in particular $\varphi'$ and $e^{4A}$ are
finite at the horizon (as usual, primes denote derivatives with
respect to $r$).  So $e^{4A} h \varphi' = 0$ at $r=r_H$.  In view of
\ScalarEOM, $e^{4A} h \varphi' = 0$ everywhere.  This is only
possibility if $\varphi$ is in fact constant.  Thus the original
singular geometry of \cite{kaloper,kachru}, supported by a free scalar
in the bulk, cannot be recovered as a limit of finite temperature
black hole solutions.  In \cite{kaloper}, an intuitive picture of a
``mechanism'' by which the four-dimensional cosmological constant
could be tuned to zero was presented in terms of a conserved current
associated with the shift symmetry $\phi \to \phi + \delta\phi$ which.
The current in some sense carries off visible sector vacuum energy
into the bulk.  This is an appealing picture, but in a geometry of the
form \AnsatzAgain\ the current has the form $J^r = e^{4A} h \partial_r
\varphi$, which vanishes at the horizon and therefore throughout the
geometry.\footnote{I thank M.~Peskin and R.~Sundrum for comments on
this point.}

We now wish to contemplate some variations on the basic Stanford
proposal, still in the general framework of \cite{vvTwo}.  First, it
is straightforward to work with any number of scalars with any bulk
potential $V(\vphi)$ which can be written in the form
  \eqn{VWFormAgain}{
   V(\vphi) = {1 \over 8} \left( {\partial W \over \partial\vphi}
    \right)^2 - {1 \over 3} W(\vphi)^2 \,.
  }
 Then a bulk geometry with $3+1$-dimensional Poincar\'e invariance can
be generated from the first order equations \FirstOrder, and it is
possible that the naked singularities that generically arise can be
obtained as limits of finite temperature black hole solutions provided
$V$ is bounded above.  Let us assume a Planck brane action of the form
  \eqn{BraneAction}{
   S_{\rm brane} = -\int d^4 \xi \, \sqrt{g_{\rm (induced)}} 
     \lambda_{\rm Pl}(\vphi) \,.
  }
 The equations of motion that determine the possible embeddings of the
Planck brane in the bulk geometry are
  \eqn{Extrinsic}{\eqalign{
   \theta_{ij} - \theta g^{\rm (induced)}_{ij} &= 
    \lambda_{\rm Pl}(\vphi) g^{\rm (induced)}_{ij}  \cr
   \partial_n \vphi &= {1 \over 2} 
     {\partial\lambda_{\rm Pl} \over \partial\vphi} \,.
  }}
 Here $\theta_{ij}$ is the extrinsic curvature of the Planck brane,
$\theta$ is its trace, $g^{\rm (induced)}_{ij}$ is the induced metric,
and $\partial_n$ is the normal derivative at the Planck brane.
Equation \Extrinsic\ admits a solution where the induced metric is
flat precisely if $\pm \lambda_{\rm Pl}(\vphi)$ is tangent to the
$W(\vphi)$ used to generate the bulk geometry.  Essentially this
condition was obtained in \cite{dfgk}.

The generalization of \cite{kaloper} is to demand that $W(\vphi) = \pm
\lambda_{\rm Pl}(\vphi)$ {\it identically} solves \VWFormAgain.  When
$V(\varphi) = 0$ for a single scalar $\varphi$, this reduces to the
special exponential $\lambda_{\rm Pl}(\varphi)$ considered explicitly
in \cite{kaloper}.  In Horava-Witten theory compactified on a
Calabi-Yau manifold, with some embedding of the spin connection in one
$E_8$, the tensions of the two ends of the universe do coincide, as
functions of the Kahler moduli, with an appropriately defined bulk
superpotential.  This seeming coincidence is a result of
supersymmetry.  The corresponding no-force condition indeed guarantees
the existence of solutions with $3+1$-dimensional Poincar\'e
invariance.  But it seems implausible that the tension of an
end-of-the-universe brane after supersymmetry breaking would solve the
equation \VWFormAgain\ identically.

The generalization of \cite{kachru} is to demand only that $W(\vphi)$
and $\pm\lambda_{\rm Pl}(\vphi)$ are tangent at some point.  This
requires exactly one fine-tuning if $W(\vphi)$ is regarded as a fixed
function of $\vphi$.  If there are $n$ scalars, then there is an
$n$-parameter family of solutions to \VWFormAgain.  If those $n$
parameters are left free, then there is an $(n-1)$-parameter family of
solutions with $3+1$-dimensional Poincar\'e invariance both in the
bulk and on the brane.  Thus one ``postpones'' the cosmological
constant problem from a single fine-tuning of the parameters of the
theory to a single fine-tuning to the actual state of the universe
(specified in this case by a choice of integration constants for
\Extrinsic).  It is not clear that there is any dynamical adjustment
mechanism to push the universe toward a flat-space solution.  (The
current mentioned after \ScalarEOM\ is only conserved in the case of a
constant potential for some scalar).

On one hand, we should note that the conjecture~\Conjecture, the
fluctuation analysis in section~\ref{Fluctuations}, and the examples in
sections~\ref{Coulomb} and~\ref{Examples} all point toward conditions
at curvature singularities which take the form of inequalities rather
than equalities.  If this is the true state of affairs, and not just
an indication of flat directions in the dual field theory or of an
insufficiently precise understanding of the microscopic physics at the
singularities, then the Stanford construction, as well as variations
of it in the spirit of \cite{vvTwo}, are probably tenable.  The only
remaining proviso is that there is not in general a clear mechanism
for preferring nearly flat solutions above all others.

On the other hand, if we believe that the bulk geometry should be the
limit of a black hole solution (which seems particularly reasonable in
a cosmological context, where the current state of the universe was
arrived at through a long cooling process from very hot initial
conditions), then the natural expectation is that boundary conditions
at the naked singularity are inherited from its near-extremal
generalizations.  For instance, in the case of a free bulk scalar, the
boundary condition in a static black hole geometry can be phrased as
zero shift current across the horizon.  Imposing this condition on
naked singularities caused by a divergence in the bulk scalar simply
rules out the existence of static singular solutions.  At best these
singularities could appear as transient states, probably with a
lifetime comparable to the five-dimensional Planck time.

The construction of \cite{BrandhuberSfetsosTwo} is an excellent
illustration of points we have argued.  The singularity is of exactly
the same type as the ones in \cite{kaloper,kachru}.  In our language,
the singularity is characterized by $\zeta = \sqrt{8/3}$.  The only
difference between \cite{BrandhuberSfetsosTwo} and
\cite{kaloper,kachru} is that in \cite{BrandhuberSfetsosTwo} there is
a potential for the bulk scalar which goes to $-\infty$ at the
singularity.  The physics is transparent if we are willing to take the
view that the bulk represents a cutoff quantum field theory, coupled
to gravity by the existence of the Planck brane \cite{gCut}.  The bulk
is precisely the five-dimensional representation of a state on the
Coulomb branch of ${\cal N}=4$ gauge theory, corresponding to a
ten-dimensional geometry where the D3-branes are arranged in an $S^3$
shell.  This is a state which seems obviously incapable of supporting
finite temperature.  Because of the bulk potential, one might hope
that the arguments following \ScalarEOM\ have no force and that there
are near-extremal generalizations.  This is in fact one of the cases
which we studied numerically (see discussion following \NFourDef),
with the conclusion that there are near-extremal geometries, but their
extremal limit is a singularity with $\zeta = \sqrt{1/6}$.  The
non-generic value of $\zeta$ amounts to having precisely one boundary
condition at the singularity, which is bad news because the problem of
fitting a flat Planck brane to the bulk geometry is once again
fine-tuned by one real parameter!

According to what was referred to the ``Weak Cosmic Censorship''
conjecture in section~\ref{Discussion}, the situation of the previous
paragraph is generic.  If weak cosmic censorship is right, then
singularities of the form \ASoln\ with $\zeta \geq \sqrt{8/3}$ cannot
be obtained as limits of regular black holes.  This seems like only a
slightly stronger statement than what we proved in
sections~\ref{FiniteT} and~\ref{Asymptotics}, but in fact $\zeta =
\sqrt{8/3}$ is both a hard case to settle and a common one in
examples.  Our evidence for ruling it out is 1) numerics on two
examples, 2) the calculations following \ScalarEOM, and 3) the field
theory intuition of the previous paragraph.  This is not an airtight
case.  However, the conjecture that singularities with $\zeta \geq
\sqrt{8/3}$ do not admit near-extremal generalizations is
mathematically well-posed and capable of proof or disproof.

From the point of view of a boundary value problem, the boundary
conditions at the Planck brane, \Extrinsic, have essentially the same
character as the boundary conditions at the true boundary of $AdS_5$
that correspond to fixing the microscopic lagrangian.  Replacing the
naked singularity either with a negative tension brane or with a black
hole horizon imposes enough boundary conditions to completely
determine the boundary value problem.  The cosmological constant is
not free, but fixed, and it still seems accidental that its value in
four-dimensional Planck units is as small as $10^{-120}$.  Going on
the AdS/CFT intuition that the naked singularity represents
interesting infrared dynamics of a quantum field theory, it seems
obvious that supersymmetry breaking should induce a positive
cosmological constant.  It would be fascinating if this intuition were
somehow wrong, and if the naked singularity could be somehow
associated with a breaking of supersymmetry that does not lead to a
four-dimensional cosmological constant.  To realize this hope in a
convincing and coherent model would require some new ideas.

Constructions such as those in \cite{kaloper,kachru}, and
near-extremal generalizations of them are nevertheless interesting.
If the Planck brane's properties can somehow be fine-tuned so that a
flat solution can be obtained, then black hole generalizations of the
Poincar\'e invariant solution allow one to obtain interesting
spatially flat FRW metrics.  The mass density above extremality of the
black hole will make a contribution to $(\dot{a}/a)^2$ in the
Friedmann equation.  This was demonstrated in detail in \cite{gCut}
for the case of an $AdS_5$ bulk, and has been considered in more
generality in \cite{Kiritsis,Kraus,Shiromizu,Binetruy,Mukohyama}.  The
black hole geometry represents hidden sector degrees of freedom at
finite temperature.

\bibliography{hades}
\bibliographystyle{ssg}

\end{document}